\documentclass[sigconf, screen, nonacm]{acmart}

\AtBeginDocument{%
  }

\usepackage{multirow}
\usepackage{xcolor}
\usepackage{stfloats}

\begin{document}

\title[Semantic Network and Idea Originality in Artificial Creativity]{Are Semantic Networks Associated with Idea Originality in Artificial Creativity? A Comparison with Human Agents}

\author{Umberto Domanti}
\affiliation{%
  \institution{Free University of Bozen-Bolzano}
  \city{Bolzano}
  \country{Italy}}
\email{umberto.domanti@student.unibz.it}

\author{Lorenzo Campidelli}
\affiliation{%
  \institution{University of Trieste}
  \city{Trieste}
  \country{Italy}}
\email{lorenzo.campidelli@phd.units.it}

\author{Sergio Agnoli}
\affiliation{%
  \institution{University of Trieste}
  \city{Trieste}
  \country{Italy}}
  \affiliation{%
    \institution{Marconi Institute for Creativity}
    \city{Bologna}
    \country{Italy}}
\email{sergio.agnoli@units.it}

\author{Antonella De Angeli}
\affiliation{%
  \institution{Free University of Bozen-Bolzano}
  \city{Bolzano}
  \country{Italy}}
  \email{antonella.deangeli@unibz.it}

\renewcommand{\shortauthors}{Domanti et al.}

\begin{abstract}

The application of generative artificial intelligence in Creativity Support Tools (CSTs) presents the challenge of interfacing two black boxes: the user's mind and the machine engine. According to Artificial Cognition, this challenge involves theories, methods, and constructs developed to study human creativity. Consistently, the paper investigated the relationship between semantic networks organisation and idea originality in Large Language Models. Data was collected by administering a set of standardised tests to ChatGPT-4o and 81 psychology students, divided into higher and lower creative individuals. The expected relationship was confirmed in the comparison between ChatGPT-4o and higher creative humans. However, despite having a more rigid network, ChatGPT-4o emerged as more original than lower creative humans. We attributed this difference to human motivational processes and model hyperparameters, advancing a research agenda for the study of artificial creativity. In conclusion, we illustrate the potential of this construct for designing and evaluating CSTs.
  
\end{abstract}



\begin{teaserfigure}
  \includegraphics[width=\textwidth]{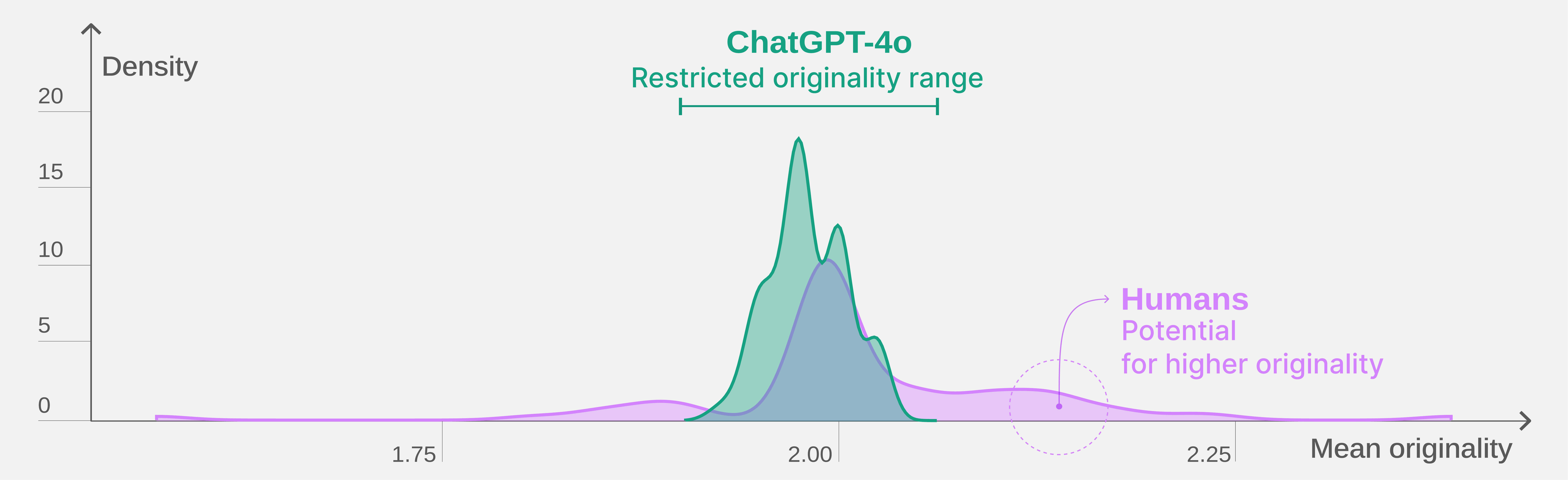}
  \caption{Kernel density estimates of the distributions of mean idea originality scores collected by administering the Alternate Uses Task to ChatGPT-4o and a sample of 81 students} 
  \Description{Figure 1 shows kernel density estimates of Alternate Uses Task mean originality scores of ChatGPT-4o and Humans. The Y-axis represents the density. It ranges from 0 to 20. The X-axis shows the mean originality. It ranges from 0 to 2.25. Two density curves are shown. ChatGPT-4o has a restricted originality range, indicating a concentrated distribution around its average and low variability. In contrast, the human curve is wider, reflecting greater variability and the potential for lower and higher originality scores.}
  \label{fig:teaser}
\end{teaserfigure}


\maketitle

 \section{Introduction}
Creativity is a complex aspect of human cognition with wide-ranging implications for individual, societal, and technological progress \cite{Agnoli_2022,Corazza_2022}. The standard psychological definition describes creativity as a multidimensional construct \cite{sternberg_investing_1996} based on the ability to produce original and effective products \cite{Runco_2012}. Despite the extensive application of this definition in many disciplines, Human-Computer Interaction (HCI) research has long lacked conceptual clarity in defining what creativity means \cite{Frich}. Furthermore, the application of generative artificial intelligence (GenAI) to creative tasks has added further terminological complexity. As a result, the word creativity is now used indifferently to denote human and machine performance, as well as their interaction \cite{geroimenko_human-computer_2025, rafner_creativity_2023,RUNCO_2023, bruno_three_2025}. 
However, psychology suggests substantial differences between human and artificial creativity in intentionality and authenticity \cite{Runco_2025, RUNCO_2023}, emotional depth and cultural understanding \cite{cropley_creations_2025}, or experience \cite{Abraham_2025}. Consistently, the study reported in this paper has highlighted  differences in the distribution of mean original ideas produced by humans and machines (Figure \ref{fig:teaser}). Human scores are distributed along a larger range of responses. On the contrary, machine scores are centred on the average performance of humans cutting out both lower and higher creative ideas ~\cite{cropley_averagecreativity_2025}.

The question of whether GenAI can be creative has stimulated a stream of  research  comparing the  performance of machines and humans using psychological tests \cite{GILHOOLY2024100071}. The results are contradictory, with some studies showing better machine performance \cite{cropley_is_2023, hubert_current_2024, GUZIK2023100065}, equal performance \cite{HAASE_2023}, or better human performance \cite{stevenson_2022putting, bangerl_2024}. Despite their  differences, this research shares an implicit product-centred definition of creativity \cite{Runco_2012}. GenAI is considered creative because it generates outcomes that would be defined creative if produced by humans \cite{Runco_2025, RUNCO_2023}. Inevitably, this definition is bound to have a limited impact on the design and development of rapidly evolving obscure models and algorithms. The product-centred definition acknowledges the occurrence of creative outputs but cannot explain how they were generated. The question is no longer whether GenAI can be creative or which model generates more creative outputs \cite{RUNCO_2023}. The pertinent question is: \emph{what are the mechanisms that could lead to artificial creativity?} Only this answer will provide designers of Creativity Support Tools, CSTs  \cite{Frich, Shneiderman, fischer_CST}, the control they need to infrastructure  effective and original interactions, which by definition are creative solutions to complex interfaces.

The distinction between human and artificial creativity \cite{RUNCO_2023, Runco_2025} is relevant to the design  of CSTs \cite{Frich, Shneiderman, fischer_CST}. The ultimate challenge is how to interface two different types of black-boxes: the human mind and nondeterministic systems. In this fussy context, the outcome of the interaction is difficult to predict \cite{vinchon_manifesto_2023}, and therefore potentially creative. Both human and artificial agents provide unique requirements for creAItive collaboration \cite{Bangerl_2025}, cyber-creativity \cite{Corazza_decalogue, corazza_cyber-creativity_2025}, or entangled authorship \cite{Bomba_2024, Bomba_2025}. 
Such complexity requires psychological knowledge \cite{Frich, Frich_CSTsHCI} because: ``Compared to nearly seven decades of pathbreaking contributions from psychological creativity research, it is evident that HCI-oriented creativity research does not have an equally strong research tradition'' ~\cite[p.3]{Frich}. To overcome this gap, the current paper investigates artificial creativity by applying theories and methods of creative cognition \cite{benedek_toward_2019}. Accordingly, creativity is the \textit{process} that emerges from the combination of basic neurocognitive mechanisms (memory, attention, and control) in divergent and convergent thinking. The paper focuses on the association between semantic memory network (the representation of the organisation of the conceptual units in memory) and idea originality in a divergent task. Psychology has demonstrated that more flexible networks are associated with higher originality \cite{kenett_network, kenett_semantic_2019}. 
Therefore, the specific research question  addressed in this paper is:
\begin{description}
    \item[RQ]  \emph{What is the relationship between semantic networks and idea originality in ChatGPT-4o}?
\end{description}

 To answer it, we conducted an experiment comparing human and artificial originality. The machine data was collected by using the chat interface of GPT-4o, which reproduced a complete black box example, as no information on model hyperparameters was available. The human sample included 81 psychology students divided into higher and lower creative individuals. Results confirmed the expected association between semantic networks and originality only with respect to higher creative individuals. Consistently, ChatGPT-4o was less original than higher creative humans,
 who also displayed a more flexible semantic memory network. However, the expected association did not manifest in the comparison with lower creative individual. Despite having a more rigid network, ChatGPT-4o emerged as more original than them.

The contribution of this paper to the CHI community is empirical, operational, and methodological. To the best of our knowledge, the paper provides the first \textit{empirical} evidence of an association between semantic networks (an indication of the creative process) and idea originality (an indication of the creative product). The relationship mirrored human creativity only with regard to the comparison with higher creative individuals. This unexpected outcome inspired new fascinating research directions for the study of artificial creativity in HCI. In short, the paper has the following organisation. Section \ref{sec:relatedWork} presents the related work about creativity in HCI and psychology, with the aim to seed a ground for our research agenda. Section \ref{sec:method} presents the method of the study advancing best psychological practices. Section \ref{sec:results} reports the results and Section \ref{sec:discussion} discusses their relevance and limitations as  directions for future work. Section \ref{sec:implications}  advances the definition of artificial creativity as a relevant construct for the design of CSTs, discusses contributions, and provides two operational insights (OPI) and three methodological recommendations (MER) for a systematic research agenda on artificial creativity. Finally, Section \ref{sec:conclusion} concludes the manuscript with a call for further research on artificial creativity.

\section{Related Work}\label{sec:relatedWork}

According to Rhodes \cite{Rhodes}, the study of creativity can be differentiated in four areas: \emph{person} (personality traits, abilities, attitudes, and self-concepts), \emph{product} (originality and effectiveness), \emph{process} (cognitive and neurophysiological mechanisms), and \emph{press} (environmental context). Following this categorisation, we identify the specific remit of HCI in the press and review research on the application of GenAI in artistic practices as a paramount example ~\cite{Chakrabarty, caramiaux, Ghajargar}. 

\subsection{HCI and Creativity}
The interest  in CSTs dates back in time \cite{Shneiderman, fischer_CST}, and extends across a wide range of  practices   \cite{Frich, Frich_CSTsHCI, Remy_CSTsHCI, ceh2024creativity}  recently challenged by the widespread adoption of AI \cite{Chakrabarty_CSTsLLM, chung_AICST_DIS2021, ning_CST_2023, bruno_three_2025}. On the one hand, GenAI provides immediate feedback and hints, thus offering users the possibility to explore new ideas \cite{Rayan, chung_AICST_DIS2021}. On the other hand, designers must be aware that these systems may distract the creative process \cite{Tankelevitch}. Negative outcomes on the long run include over-reliance and homogenisation \cite{doshi_generative_nodate, Kumar, Anderson, bruno_three_2025, Bangerl_2025}. In artistic domains,  HCI researchers and designers should take into account that GenAI may eventually lead to a reduction of variability and diversity in divergent thinking  \cite{Kumar, doshi_generative_nodate, Anderson}, and induce a form of Creative Mortification \cite{beghetto_creative_2014} where people perceive their own work as less valuable than those generated by AI. 
Additional concerns relate to ownership, plagiarism, and originality \cite{Rajcic, Chang, vinchon_manifesto_2023}. Unresolved constraints have emerged in the context of creative writing due to the inability of commercial LLMs to deal with conflict or humour, which are the essence of tragedy and comedy, respectively \cite{Grigis_2024}.

Despite its implicit limitations, some artists recognise a creative potential in GenAI \cite{caramiaux, Bomba_2024, Bomba_2025}. Particularly, they found hints of inspirations in its ``hallucinations'' \cite{Bomba_2024, caramiaux} --- the generation of plausible information that is factually incorrect or unsupported by data \cite{geroimenko_hall_2025}. Following the standard definition of creativity \cite{Runco_2012}, we associate this preference with the intrinsic originality of hallucinations. Far from interpreting them as system bugs, artists value hallucinations as seed incidents, surprising and intriguing clues that can inspire their work \cite{deangeli_2025, rastelli_hallucinations, Bomba_2024}.
Altogether, this research demonstrates that occasionally GenAI produces outcomes that are perceived as creative \cite{RUNCO_2023}. A provocative example is Théâtre D’opéra Spatial, a ``painting'' generated by Midjourney following a prompt by Jason M. Allen. The artwork won the fine art competition at the 2022 Colorado State Fair. Yet, the U.S. Registration Office denied copyright to the artist based on the impossibility of differentiating between human and artificial contributions in the creative process \cite{Bomba_2025}.

\subsection{Creativity as a Cognitive Process}

To move beyond product-oriented research \cite{Runco_2012}, we follow the definition of creativity as the dynamic interplay between inconclusiveness (when originality or effectiveness has not yet been reached) and achievement (the fulfilment of both originality and effectiveness). In other words, creativity requires ``potential'' originality and effectiveness \cite{Corazza_2016}. The term potential allows the shift from a focus on the product to a focus on the process. It acknowledges the existence of countless instances in which the value of a product was not initially understood but was rediscovered and recognised later, as well as cases in which the outcome itself is not primarily important \cite{corazza_cyber-creativity_2025}. 

\begin{figure}[t]
    \centering
    \includegraphics[width=\linewidth]{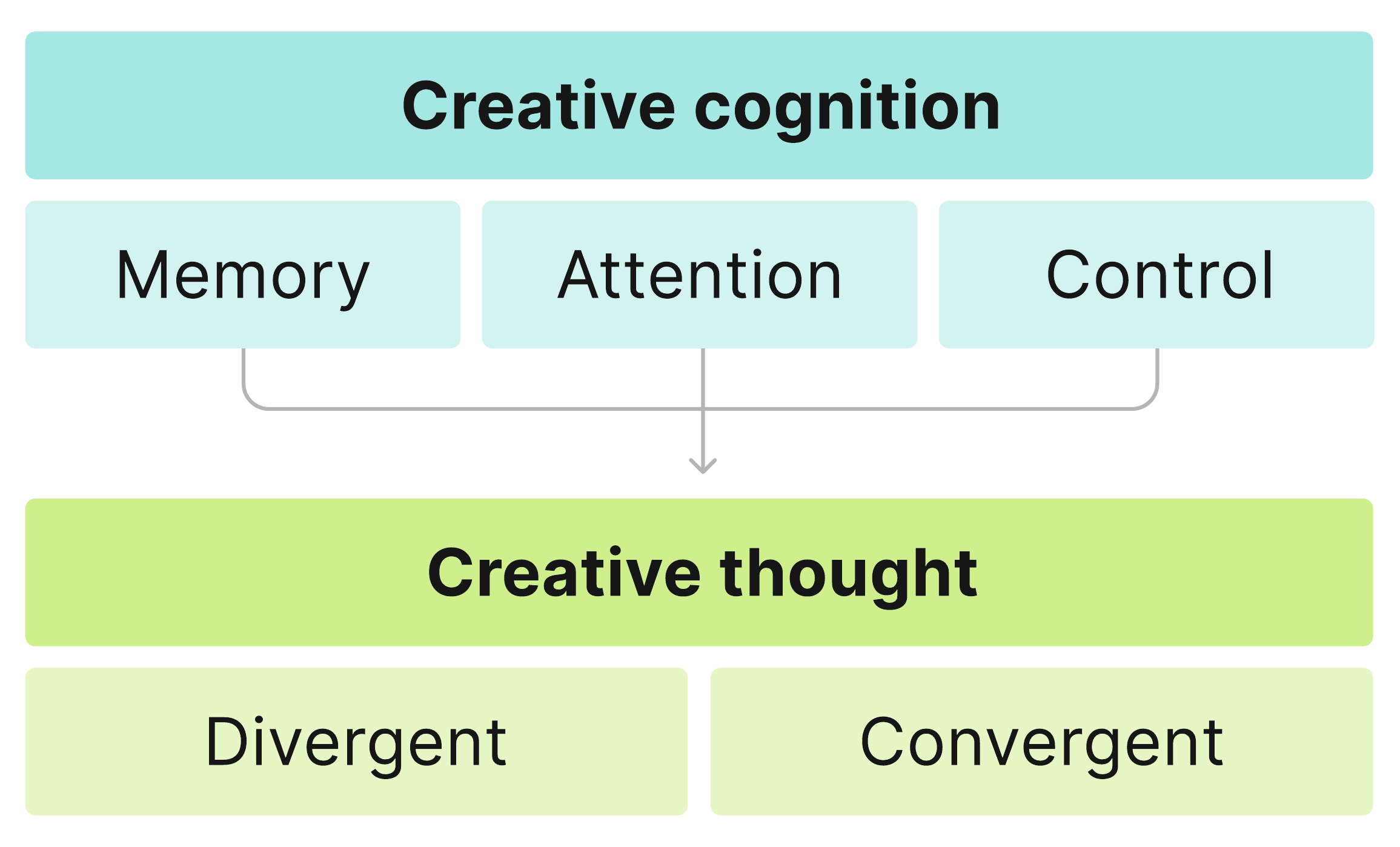}
   \caption{Creative Cognition framework, adapted by \cite{benedek_toward_2019}} 
   \Description{Figure 2 illustrates the creative cognition framework. At the top of the diagram, there is a box with the writing Creative cognition. Under this box, three smaller boxes are arranged horizontally: on the left memory, in the middle attention, and on the right control. There is an arrow starting from these three boxes and leading to a lower box with the writing Creative thought. Under it, two smaller boxes: on the left divergent and on the right convergent}
    \label{fig:Creative Cognition}
\end{figure}

As illustrated in Figure \ref{fig:Creative Cognition}, the creative process emerges from basic neurocognitive abilities: memory, attention, and cognitive control \cite{benedek_toward_2019}. They converge into two generative modalities: convergent and divergent thinking \cite{GUILFORD}. Convergent thinking considers all available information to assemble it in the most compact form \cite{gabora2018neural} and possibly arrive at a single result through integration \cite{lubart2013creative}. In contrast, divergent thinking is an exploratory cognitive mode that aims to generate multiple solutions or ideas for a problem or task. The current paper investigates the relationship between semantic memory network as a basic mechanism of the creative process~and idea originality as an indicator of divergent creative outcomes. 

\subsection{Divergent Creativity}

Previous works have demonstrated how the divergent creative potential of LLMs is improving over time \cite{Vinchon}. Furthermore, research suggests that LLMs reflect typical human patterns in idea generation, such as the serial order effect, whereby later responses tend to be more novel than earlier ones \cite{beaty_why_2012,GILHOOLY2024100071, stevenson_2022putting}. Other authors have suggested that it is possible to enhance LLM creativity by incrementally prompting the model more forcefully to generate original responses across interactions \cite{goes_2023pushing}. Most of these studies investigated artificial creativity using experimental divergent tests.

 As a reliable indicator of real world achievements \cite{Jauk, said-metwaly_divergent_2024}, divergent thinking can be assessed through a set of  tasks, including verbal and visual stimuli \cite[TTCT;][]{torrance_1966}, ambiguous figures \cite[FIQ;][]{grassini_artificial_2025} and the production of words that are as different from each other as possible \cite[DAT;][]{olson_naming_2021}. A common verbal test is the Alternate Uses Task \cite [AUT;][]{GUILFORD}, which requires generating as many alternative uses as possible for common objects (e.g. a brick) within a fixed time. Responses provide a measure of fluency (the number of generated ideas), flexibility (the number of different categories of ideas), elaboration (the level of idea detail), and originality (the idea novelty and uncommonness). Historically, originality has been assessed based on interrater agreement between at least two expert judges on a Likert scale \cite{agnoli_dopamine_2023}. Recently, however, deep neural network-based LLMs have been trained to perform this task. An example is OCSAI (Open creativity scoring with artificial intelligence), which achieved a correlation $r = 0.81$ with human raters \cite{organisciak_OCSAI}.

  \begin{table*}
  \caption{Overview of studies comparing humans and LLMs with divergent creativity tasks}
  \label{tab:humans_llms_creativity}
  \Description{Table 1 summarises the eight published studies comparing human participants and Large Language Models performances. For each study, the table provides details on the comparison between humans and Large Language Models, the creativity assessment, the main limitations and the main findings.}
\begin{tabular}{p{0.8cm} p{4cm} p{4cm} p{4cm} p{3cm}}
    \toprule
    \textbf{Paper} & \textbf{Comparison} & \textbf{Creativity Assessment} & \textbf{Main \newline Limitations} & \textbf{Main \newline Findings} \\
    \midrule
\cite{stevenson_2022putting} & 42 psychology students vs ChatGPT-3 on AUT. & 1. Two human coders rated originality, utility, and surprise.\newline 2. Semantic distance.\newline 3. Flexibility. & 1. Small human sample. & H $>$ LLM in originality, surprise, flexibility, semantic distance. \newline LLM $>$ H in utility.\\
\midrule
\cite{bangerl_2024} & 20 psychology students in couples vs 10 chatbots on AUT. & 1. Fluency. \newline 2. Flexibility. \newline 3. Originality. \newline 4. Elaboration. & 1. Small human sample. & H $>$ LLMs \newline in originality, flexibility. LLMs $>$ H \newline in elaboration. \\
\midrule
\cite{HAASE_2023} & 100 participants vs 6 chatbots (Alpa.ai, Copy.ai, ChatGPT-3/4, Studio.ai, YouChat) on AUT. & 1. Six human coders and OCSAI rated originality. & 1. Paid Prolific participants. & LLMs = H. \newline 9\% of H $>$ LLMs. \\ 
\midrule
\cite{koivisto_best_2023} & 256 participants vs 3 chatbots (ChatGPT-3/4, Copy.ai) on AUT. & 1. Semantic distance. \newline 2. Six trained judges rated originality. & 1. Paid Prolific participants. \newline 2. Small number of chatbot observations. & LLMs $>$ H. \newline Best H = or $>$ LLMs. \\ 
\midrule
\cite{grassini_artificial_2025} & 256 participants \newline vs ChatGPT-4 on FIQ. & 1. Flexibility (semantic distance between interpretations). \newline 2. Subjective Perceived Creativity. & 1. Paid Prolific participants. \newline 2. Only four \newline figures analysed. \newline 3. Fluency and elaboration controlled. & LLM $>$ H in flexibility. H $>$ LLM in subjective creativity. \newline Best H $>$ LLM \newline in flexibility and subjective creativity. \\
\midrule
\cite{GUZIK2023100065}& 24 undergraduates \newline + school performance data vs ChatGPT-4 on TTCT. & 1. TTCT standard scoring (fluency, flexibility, originality) by human scorers \newline at STS. & 1. Small human sample. 2. Only one TTCT task. & LLM $>$ H \newline in fluency, flexibility, and originality. \\
\midrule
\cite{hubert_current_2024} & 151 participants vs GPT-4 on AUT, DAT, TTCT. & 1. Fluency. \newline 2. OCSAI rated originality and elaboration. & 1. Paid Prolific participants. \newline 2.Only one TTCT task. & LLM $>$ H in originality and elaboration. \\ 
\midrule
\cite{cropley_is_2023} & Large human baseline \cite{olson_naming_2021} vs ChatGPT-3/4 on DAT. & 1. Semantic distance. & 1. Small differences in mean scores between groups. & LLMs $>$ H; \newline Concerns on AI reliability. \\ \bottomrule

\end{tabular}
\end{table*}

The findings of experimental studies comparing humans and LLMs on divergent thinking tasks are contradictory, likely reflecting methodological and procedural differences, each with its own strengths and limitations. Table \ref{tab:humans_llms_creativity} provides an overview starting from the articles reviewed by Gilhooly \cite{GILHOOLY2024100071} and including only published papers which underwent peer-review by July 2025. Most of these studies addressed originality, with a preference for the AUT. Based on such comparison, some experiments have shown that humans outperform machines \cite{bangerl_2024,stevenson_2022putting}, while others have demonstrated that machines have higher divergent thinking capabilities than humans \cite{cropley_is_2023, GUZIK2023100065, hubert_current_2024}. The inconsistency can be explained by several factors. These include procedural differences in the collection of human and machine data; cognitive, affective, and motivational variations in the human sample; and the evolving capabilities of LLMs. Furthermore, at a conceptual level, the comparison between humans and machines is oversimplified.

 As reported in Table \ref{tab:humans_llms_creativity}, some studies revealed that the machine performance outperformed human average, but others noted how a small portion of humans (the higher creative individuals) consistently outperformed the machine \cite{koivisto_best_2023, HAASE_2023, grassini_artificial_2025}. Therefore, this paper adopted an experimental design that includes Higher Creative Humans (HCH) and Lower Creative Humans (LCH). Within this methodology, we posit that HCH will outperform ChatGPT-4o (GPT) in originality, while GPT is expected to be more original than LCH (\emph{H1.1}).

\subsection{Semantic Memory}

Memory plays a crucial role in human creativity \cite{benedek_toward_2019, benedek_role_2023}. However, the study of the memory-creativity link is complex as there are different types of memory (working, semantic, episodic, short/long-term) and different modalities of creative thinking (e.g., divergent and convergent). According to the MemiC framework, the role of memory in the creative ideation stages extends from the search for memory and the construction of candidate ideas to the evaluation of novelty and effectiveness ~\cite{benedek_role_2023}. In particular, the role of semantic memory in creative thinking has recently been explored. Recent research has highlighted how semantic memory structures and the ability to access these structures are fundamentally related to creative performance across the lifespan \cite{COSGROVE2021104631, cosgrove_age_2023, cosgrove_associative_nodate, campidelli_creativity_2026}.

The structure of semantic memory can be represented as a network and analysed based on mathematical graph theory \cite{Siew}. The network consists of nodes (the basic unit of the system) and edges (the semantic associations between two concepts). Data to construct these networks is generally collected by verbal production tasks \cite{kenett_network, Li03072021}. They require participants to generate as many words as possible belonging to a given semantic category or related to a target word, within a fixed amount of time. In the resulting network, each node represents a category exemplar (e.g., tiger), and the edges represent associations between two exemplars. In other words, these associations constitute the tendency to generate an exemplar \emph{b} (lion) when \emph{a} (tiger) is produced.

A large body of psychological research has shown how the semantic network of higher creative individuals differs from that of lower creative individuals \cite{benedek_how_2017,he_relation_2020, kenett_network, kenett_semantic_2019, ovando-tellez_brain_nodate}. As summarised in Table ~ \ref{tab:network_indices}, the former is characterised by (a) lower distance to traverse from one node to another, ASPL; (b) higher connectivity between conceptual nodes, CC; and (c) fewer sub-communities within the network, Q \cite{kenett_network, kenett_semantic_2019}. Moreover, higher creative individuals appear to have a more resilient and flexible network, which is more resistant to targeted “attacks” than lower creative individuals, R \cite{Kenett_percolation}. Thus, a robust and flexible structure enables the activation of a broader range of associations, thereby increasing the possibility of generating new ideas \cite{Kenett_percolation}. 

Despite semantic network flexibility being a reliable indicator of human creative performance, it remains unclear how such structures are represented in LLMs \cite{cropley_creations_2025, koivisto_best_2023}. Moreover, much uncertainty exists about the association between network organisation and idea originality in artificial creativity. To the best of our knowledge, only one published study has investigated the difference in semantic network structures between humans and different LLMs \cite{WANG2025100103}. Human data was obtained from a random selection of 30 participants in the animal category of the semantic fluency database reported in \cite{zemla_snafu_2020}. LLM data was extracted by instructing the LLMs to play specific roles related to 30 common occupations (e.g., pretend that you are a chef) before performing the task. This procedure was intended to enrich the diversity of the machine responses \cite{shanahan_role_2023}. The results showed that the semantic network of LLMs has higher ASPL, lower CC, and higher Q compared to humans \cite{WANG2025100103}. Despite creativity not being a variable of the study, the authors concluded that ``LLMs have poorer creativity and flexibility compared to humans''~\cite[p. ~6]{WANG2025100103}. 

The current study was designed to explicitly study the association between semantic networks and originality in artificial vs. human creativity. Following Artificial Cognition \cite{taylor_artificial_2021} and expanding the study of Wang et al. \cite{WANG2025100103}, we expected GPT to have significantly different semantic networks than HCH and LCH, given the expected direction of idea originality stated in \emph{H1.1.} We tested this hypothesis with structural and percolation analysis using the measures described in Table \ref{tab:network_indices}. Consistently, we hypothesise that GPT would display a more flexible network compared to LCH and the reverse compared to HCH (\emph{H1.2}).  

  \begin{table*}
  \caption{The network measures used to test \emph{H1.2}: Network Flexibility}
   \Description{Table 2 presents the description of the structural network measures (Average Short Path Length, Clustering Coefficient, Modularity) and the percolation network measure (Resilience) used to test hypothesis 1.2 related to network flexibility. For each measure, the table shows the expected direction of group differences.} 
  \label{tab:network_indices}
  \begin{tabular}{{l p{2cm} l p{8.5cm} l}}
    \toprule
    \textbf{Analysis} &\textbf{Measure} & \textbf{Acronym} & \textbf{Description} & \textbf{Expected Results } \\
    \midrule
    & Average Short Path Length & ASPL & Average shortest number of steps (i.e., edges) needed to traverse between any pair of nodes. A higher value means the network is more spread out. & LCH $>$ GPT $>$ HCH \\
   Structural & Connectivity & CC & Clustering coefficient of a network, describing how much the neighbours of a node are interconnected. A higher value means more interconnection. & LCH $<$ GPT $<$ HCH \\
    & Modularity & Q & Degree to which a network can be divided into sub-communities. A higher value means more segregation. &LCH $>$ GPT $>$ HCH \\
    \midrule
    Percolation &Resiliency & R & Measure of flexibility of a network. A higher value means more resilience. & LCH $<$ GPT $<$ HCH \\
    \bottomrule
  \end{tabular}
\end{table*} 

\section{Method}\label{sec:method}

The study was designed to test the hypothesis that the semantic network flexibility of an LLM provides an indication of its idea originality \emph{(H1)}. Instantiating the principle of Artificial Cognition \cite{taylor_artificial_2021} with research on human creativity \cite{benedek_toward_2019, WANG2025100103}, we refined it into two sub-hypotheses, both of which were expected to be true.

\begin{itemize}
    \item \emph{H1.1} --- Idea Originality: HCH $>$ GPT $>$ LCH
    \item \emph{H1.2} --- Network Flexibility: HCH $>$ GPT $>$ LCH
\end{itemize}

\subsection{Participants and Design}\label{sec:participants}

Human responses were obtained from the dataset collected by Campidelli et al. \cite{campidelli_creativity_2026}. The sample analysed in the present study included 81 psychology students (mean age = 20.3 years, SD = ~1.71, men = 12, women = 69). Machine data was obtained on the 6th and 7th of March 2025 by using ChatGPT-4o chat interface\footnote{https://chat.openai.com} with the ``Plus'' subscription package \cite{Vinchon}. This interface neither displayed nor allowed for the modification of the values of the hyperparameters. Based on the mean originality scores of each human participant, the human sample was median split into (a) LCH and (b)  HCH. Accordingly, the experiment compared the networks and creative performances of three groups: GPT, LCH, and HCH.

\subsection{Procedure}
The experiment with human participants took place in a room free of acoustic and visual distractors and was conducted by an expert researcher. After signing informed consent, they sat in front of an empty desk and were asked to complete two different verbal production tasks and the AUT. The target words were printed on a sheet of paper, which remained on the desk for the entire duration of the tasks. The responses were recorded with an audio recorder and manually transcribed by one of the experimenters. The procedure was repeated by creating a new separate chat for each of the 81 ChatGPT-4o participants (Appendix \ref{sec: appendixA}) \cite{Vinchon}. All human and machine responses were merged, anonymised,  sorted in alphabetic order within each object, and finally scored for originality by two expert coders, blind to the experimental goals and the origin of the data, as recommended by Silvia et al. \cite{silvia_assessing_2008}. The interrater reliability was high, ICC = .93.

\subsection{Experimental Tasks}\label{sec:tasks}
Appendix \ref{sec: appendixA} presents the instructions given to humans and prompts received by ChatGPT-4o across three tasks. The Verbal Fluency Test and the Free Association Task were used to construct semantic networks. Finally, the AUT was used to estimate originality. The Verbal Fluency Test required participants to produce as many words as possible belonging to the ``animals'' and ``fruit and vegetables'' categories within one minute for each target \cite{goni_semantic_2011}. The Free Association Task required participants to verbally produce as many words as possible associated with ``building'' and ``notebook'' within one minute for each target \cite{nelson_what_2000}. The presentation order of the verbal production tasks was randomised. For the AUT, human participants were asked to generate as many original alternative uses as possible for ``shoe'' and ``fork'' within 3 minutes for each target.  ChatGPT-4o was prompted with the same instruction provided to human participants (Appendix \ref{sec: appendixA}). Appendix \ref{sec: appendixB} provides examples of interactions (i.e.,  prompts and outputs), for both humans and ChatGPT-4o. The databases containing all  responses for each task and the scripts for the analyses are available in the Supplementary Materials. 

\subsection{Semantic Networks}\label{subsec: semantic memoy network}
The network estimation followed the procedure and pipelines provided by Christensen et al. \cite{christensen_tutorial2023}. The entire verbal production database was translated from Italian to English and pre-processed with the R packages \emph{SemNetDictionaries} and \emph{SemNetCleaner}. The cleaned data was entered into a binary response matrix with participants in rows and words in columns. Each cell was filled with 1 if the participant had produced the word and 0 if they had not. The networks were estimated using the \emph{SemNet} package and applying the correlation-based method \cite{christensen_tutorial2023}. The cosine similarity function was employed to estimate the edges between nodes and compute the association profiles of the responses. To control for confounding factors (e.g., different numbers of nodes or edges among groups), the matrices only included responses that were provided by at least two participants in each group \cite [see, for example,][]{BORODKIN, COSGROVE2021104631}. Furthermore, in line with the literature on network construction \cite{christensen_network_2018, christensen_tutorial2023, WANG2025100103, COSGROVE2021104631,Li03072021, massara_network_2016, christensen_remotely_2018}, the Triangulated Maximally Filtered Graph (TMFG) was adopted to minimise noise and spurious association by creating different networks with the same number of nodes and edges \cite{christensen_tutorial2023, christensen_network_2018, massara_network_2016}. This method aims to maximise the strength of node associations while maintaining the extant network planar to be depicted graphically \cite{christensen_network_2018}. TMFG represents the only consolidate alternative to the  Planar Maximally Filtered Graph, PMFG \cite{kenett_PMFG_2013, tumminello_tool_2005, massara_network_2016}. Generally, the two methods align with one another \cite{christensen_tutorial2023}, but TMFG is more efficient than PMFG \cite{massara_network_2016}. The resulting networks of GPT, LCH, and HCH were compared using structural and percolation analyses.

\begin{figure}[b]
    \centering
\includegraphics[width=\linewidth]{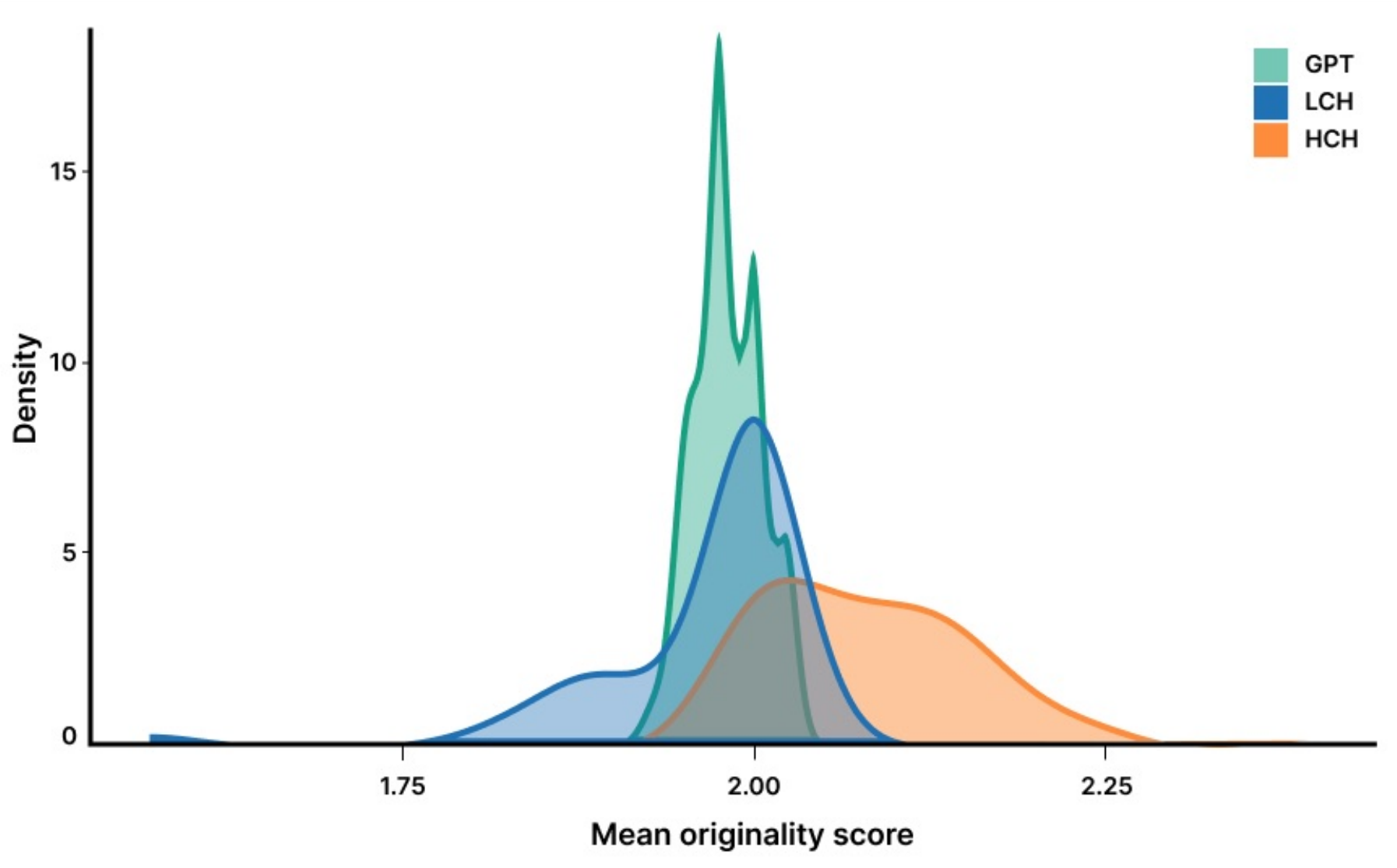}
   \caption{Kernel density estimates of mean originality scores across groups}
   \Description{Figure 3 shows kernel density estimates of Alternate Uses Task mean originality scores of ChatGPT-4o, Lower Creative Humans, and Higher Creative Humans. The Y-axis represents the density. It ranges from 0 to 15. The X-axis shows the mean originality. It ranges from 0 to 2.25. Three density curves are shown. ChatGPT-4o curve is narrow, indicating a concentrated distribution around its average and low variability. Lower Creative Humans curve partially overlaps with the one of ChatGPT-4o, but it also includes lower mean originality scores. Higher Creative Humans curve partially overlaps with the one of ChatGPT-4o, but it also includes higher mean originality scores.}
    \label{fig:density}
\end{figure}

\section{Results}\label{sec:results}

Figure \ref{fig:density} illustrates the kernel density estimates\footnote{A kernel density estimate provides a smooth curve that represents the data distribution} of AUT mean originality across groups. The  distribution of the human scores allowed us to differentiate between HCH and LCH by median split for the measure of total production and originality peaks. In contrast, the distribution of GPT scores showed a more restricted range. Its peak overlapped with the human average with minimal standard deviation, which did not allow the median split of the machine performance into different levels.

Table \ref{tab:descriptive_stats}  presents descriptive statistics of fluency for the three groups. An ANOVA revealed a significant difference,  $F_{(2,159)} = 438$, $p < .001$, $\eta^2p = 0.846$. The mean fluency of GPT was more than double in comparison to the fluency emerged in both HCH, $t_{(159)} = 26.8$, $p < .001$, and LCH, $t_{(159)} = 20.74$, $p < .001$. Finally, HCH showed higher mean fluency than LCH, $t_{(159)} = 5.08$, $p < .001$.

\begin{table}[t]
  \caption{Descriptive statistics of fluency by group}
\Description{Table 3 shows descriptive statistics of fluency scores for the three groups} 
  \label{tab:descriptive_stats}
  \begin{tabular}{l l c c c c c}
    \toprule
    \textbf{Measure} & \textbf{Group} & \textbf{Mean} & \textbf{Median} & \textbf{SD} & \textbf{Min} & \textbf{Max} \\
    \midrule
         & GPT         & 43.8 & 40   & 5.36  & 35   & 55   \\
    Fluency            & LCH        & 12.2 & 11   & 6.87  & 2    & 31   \\
                & HCH        & 19.1 & 19   & 7.03  & 8    & 36   \\
    \bottomrule
  \end{tabular}
\end{table}

\setcounter{table}{4}
\begin{table*}[b]
  \caption{Comparisons of network measures across groups}
 \Description{Table 5 reports statistical comparisons of three structural network measures across the three groups. For each structural measure, the table shows pairwise group comparisons, including degrees of freedom, test statistics, p-value, Cohen's d for effect size, mean and standard deviation of group 1, mean and standard deviation of group 2, and direction of differences.}
 \label{tab:comparisons}
  \begin{tabular}{p{1cm} p{2cm} c r c c r r r r l}
    \toprule
    \textbf{Measure} & \textbf{Comparison} & \textbf{df} & \textbf{t-statistic} & \textbf{p-value} & \textbf{d} & \textbf{Mean1} & \textbf{SD1} & \textbf{Mean2} & \textbf{SD2} & \textbf{Direction} \\
    \midrule
     & GPT vs LCH & 1998 & 51.91 & < .001 & 2.32 & 4.378 & 0.53 & 3.429 & 0.24 & GPT $>$ LCH \\
     ASPL    & GPT vs HCH & 1998 & 56.93 & < .001 & 2.55 & 4.378 & 0.53 & 3.325 & 0.26 & GPT $>$ HCH \\
         & LCH vs HCH & 1998 & 9.38  & < .001 & 0.42 & 3.429 & 0.24 & 3.325 & 0.26 & LCH $>$ HCH \\
    \midrule
       & GPT vs LCH & 1998 & -8.76 & < .001 & 0.39 & 0.705 & 0.01 & 0.710 & 0.01 & GPT $<$ LCH \\
      CC   & GPT vs HCH & 1998 & -19.84 & < .001 & 0.89 & 0.705 & 0.01 & 0.715 & 0.01 & GPT $<$ HCH \\
         & LCH vs HCH & 1998 & -9.15 & < .001 & 0.41 & 0.710 & 0.01 & 0.715 & 0.01 & LCH $<$ HCH \\
    \midrule
        & GPT vs LCH & 1998 & 31.42 & < .001 & 1.41 & 0.692 & 0.01 & 0.666 & 0.02 & GPT $>$ LCH \\
     Q    & GPT vs HCH & 1998 & 37.17 & < .001 & 1.66 & 0.692 & 0.01 & 0.662 & 0.02 & GPT $>$ HCH \\
         & LCH vs HCH & 1998 & 4.40  & < .001 & 0.20 & 0.666 & 0.02 & 0.662 & 0.02 & LCH $>$ HCH \\
    \bottomrule
  \end{tabular}
\end{table*}

\setcounter{table}{3}
\begin{table}[H]
\caption{Descriptive statistics of total production and originality peaks by group}
\Description{Table 4 reports descriptive statistics for two measures (total production and originality peaks) across the three groups}
\label{tab:originality}
\begin{tabular}{p{1.5cm}lccccc}
\toprule
\textbf{Measure} & \textbf{Group} & \textbf{Mean} & \textbf{Median} & \textbf{SD} & \textbf{Min} & \textbf{Max} \\
\midrule
\multirow{3}{*}{\shortstack[l]{Total\\production}} 
                        & GPT & 1.99 & 1.98 & 0.03 & 1.93 & 2.07 \\
                        & LCH & 1.97 & 2.00 & 0.08 & 1.57 & 2.00 \\
                        & HCH & 2.11 & 2.10 & 0.08 & 2.02 & 2.44 \\
    \midrule                                      
\multirow{3}{*}{\shortstack[l]{Originality\\peaks}} 
                        & GPT & 2.25 & 2.17 & 0.32 & 2.00 & 3.50 \\
                        & LCH & 2.00 & 2.00 & 0.00 & 2.00 & 2.00 \\
                        & HCH & 2.59 & 2.42 & 0.46 & 2.17 & 4.00 \\
\bottomrule

\end{tabular}
\end{table}

\subsection{Originality}\label{originality}
 We tested \emph{H1.1} related to idea originality with two linear mixed models. The first one included the originality of all AUT responses generated by each human and machine participant (total production). The second one considered only the top 3 original responses generated by each participant (originality peaks). Descriptive statistics are reported in Table \ref{tab:originality}.

\subsubsection{Total Production}
To investigate differences in originality among the three groups, we run a linear mixed model with the lme4 package \cite{bates_fitting_2015} and lmerTest \cite{kuznetsova_lmertest_2017} in R. The analysis included the entire set of responses as the dependent variable. The fixed effect was Group (GPT, LCH, HCH), while intercepts for participants’ ID and object (shoe and fork) served as random effects. As hypothesised, results showed that GPT originality scores were significantly higher than those of LCH, $B = 0.03$, $SE = 0.013$, 95\% $CI$ $[0.0003, 0.052]$, $t_{(359.46)} = 1.98$, $p = .048$, and lower than those of HCH, $B = - 0.12$, $SE = 0.011$, 95\% $CI$ $[-0.144, -0.102]$, $t_{(193.59)} = -11.35$, $p < .001$. Moreover, HCH originality scores were significantly higher than those of LCH, $B = 0.15$, $SE = 0.016$, 95\% $CI$ $[0.119, 0.180]$, $t_{(391.45)} = 9.51$, $p < .001$.

\subsubsection{Originality Peaks}
In this analysis, the top 3 original responses per participant were included as the dependent variable. The fixed effect was Group (GPT, LCH, HCH), and intercepts for participants’ ID served as a random effect. Results showed that GPT originality scores were significantly higher than those of LCH,  $B = 0.24$, $SE = 0.07$, 95\% $CI$ $[0.11, 0.38]$, $t_{(160.06)} = 3.59$, $p < .001$, and lower than those of HCH, $B = -0.34$, $SE = 0.06$, 95\% $CI$ $[-0.46, -0.22]$, $t_{(158.95)} = -5.51$, $p < .001$. Finally, HCH originality scores were significantly higher than those of LCH, $B = 0.59$, $SE = 0.08$, 95\% $CI$ $[0.44, 0.74]$, $t_{(159.85)} = 7.77$, $p < .001$.

\subsection{Semantic Networks}\label{Semantic network results}

\begin{figure*}
    \centering
    \includegraphics[width=\linewidth,trim=0cm 2cm 0cm 2.8cm,clip]{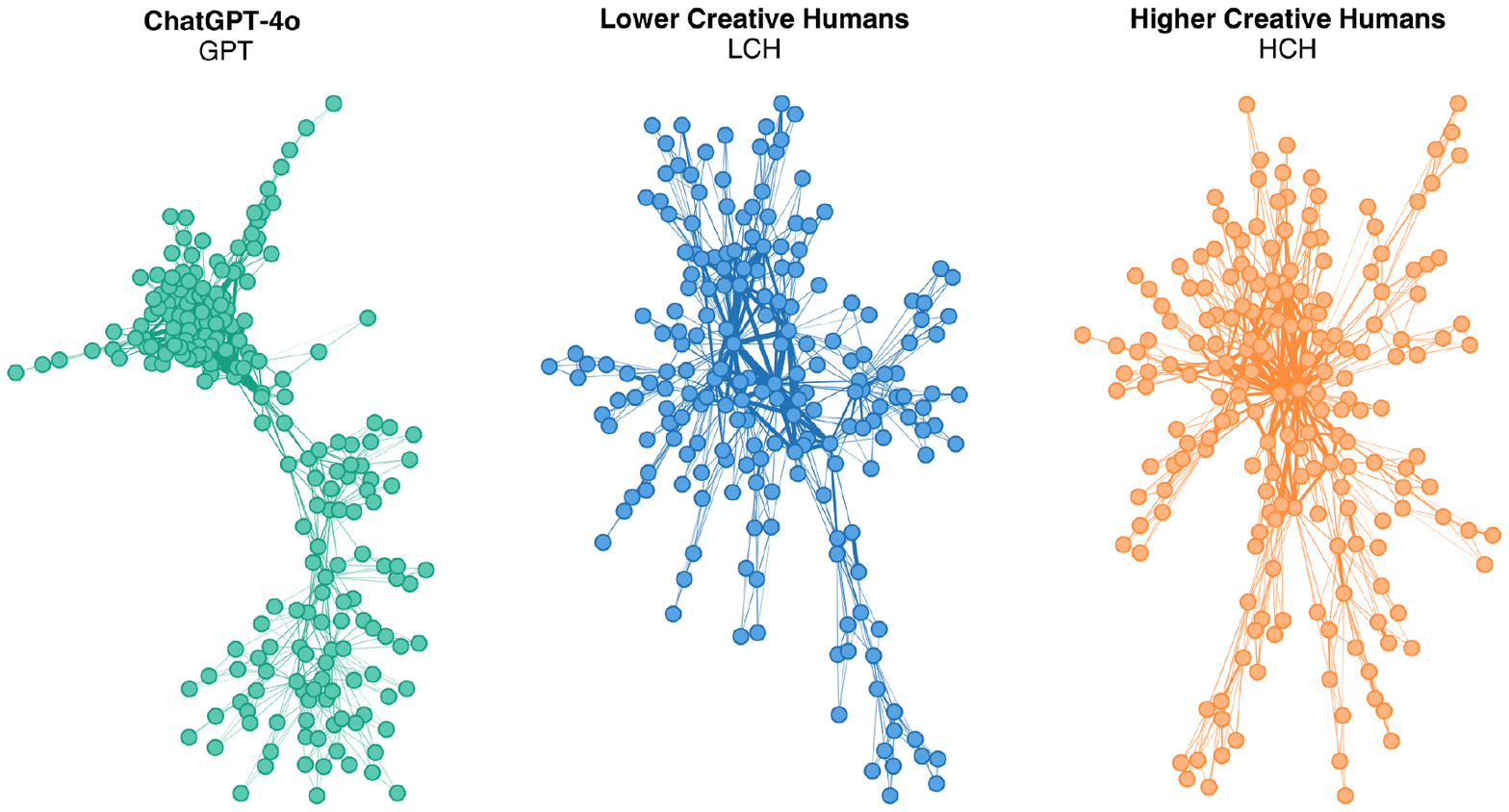}
    \caption{2D representation of GPT, LCH, and HCH semantic networks}
    \Description{Figure 4 illustrates ChatGPT-4o, Lower Creative Humans, and Higher Creative Humans semantic networks. On the left, there is the ChatGPT-4o network. In the middle, there is the Lower Creative Humans network. On the right, there is the Higher Creative Humans network. ChatGPT-4o network is more segregated and less interconnected compared to the other two human networks. The Higher Creative Humans network is more interconnected than that of Lower Creative Humans.}
    \label{fig:networks}
\end{figure*}

Figure \ref{fig:networks} illustrates a 2D graphical representation of GPT, LCH, and HCH semantic networks. There were apparent visual differences that were  compared by structural and percolation analyses.

\subsubsection{Structural Results}
Bootstrapping \cite{b_efron_bootstrap_1979} was used to estimate the network measures detailed in Table \ref{tab:network_indices}. The node-wise approach was run on 50\% of the data with 1000 iterations, assuming that if the full networks differ from each other, then the partial network consisting of the same nodes should also differ \cite{bertail_second-order_1997,politis_large_1994,Shao}. This method was adopted because it is the most conservative of the node-wise approaches, aiming at reporting only significant differences. The measures ASPL, CC and Q were computed for each group and compared with an ANCOVA with Group as independent variable and the number of edges as a covariate \cite{christensen_tutorial2023}. Post hoc comparisons were performed applying the Bonferroni correction. Results are described below and summarised in Table \ref{tab:comparisons}.

\begin{itemize}

\item \emph{Average Short Path Length (ASPL)}: The analyses revealed a significant main effect of Group, $F_{(2, 2996)} = 2430.562$, $p < .001$, $\eta_p^2 = 0.619$. Post hoc comparisons revealed that GPT showed significantly higher ASPL compared to LCH and HCH. Moreover, LCH showed higher ASPL in comparison to HCH.

\item\emph{Clustering Coefficient (CC)}:
The analyses revealed a significant main effect of Group,
$F_{(2, 2996)} = 352.412$, $p < .001$, $\eta_p^2 = 0.19$. Post hoc comparisons revealed that GPT showed significantly lower CC compared to LCH and HCH. Furthermore, LCH showed lower CC in comparison to HCH.

\item\emph{Modularity (Q)}:
The analyses returned a significant main effect of Group, $F_{(2, 2996)} = 942.253$, $p < .001$, $\eta_p^2 = 0.386$. Post hoc comparisons revealed that GPT showed significantly higher Q compared to LCH and HCH. Finally, LCH showed higher CC in comparison to HCH.
\end{itemize}

\subsubsection{Percolation Results}
Percolation analysis investigates the flexibility of a network under targeted attacks \cite{farkas_weighted_2007,palla_uncovering_2005} by removing edges that are below a given threshold for each percolation step \cite{Kenett_percolation}. The analysis was conducted using the \emph{CliquePercolation} package in R \cite{COSGROVE2021104631}. The size of the largest cluster of connected nodes (Largest Connected Component, LCCS) was measured at any step. The intensity threshold (\emph{I}, Min 0, Max 1) was gradually increased by a coefficient of 0.01. This causes the networks to transition from being totally intact (\emph{I} = 0) to maximally fragmented (\emph{I} = 1). Finally, the integral of percolation (i.e., the area under the curve) for each group was calculated considering the number of node communities in the semantic network across a range of threshold values between 0.01 and 1 \cite{COSGROVE2021104631}. 
\begin{figure*}[!b]
        \centering
\includegraphics[width=0.85\linewidth]{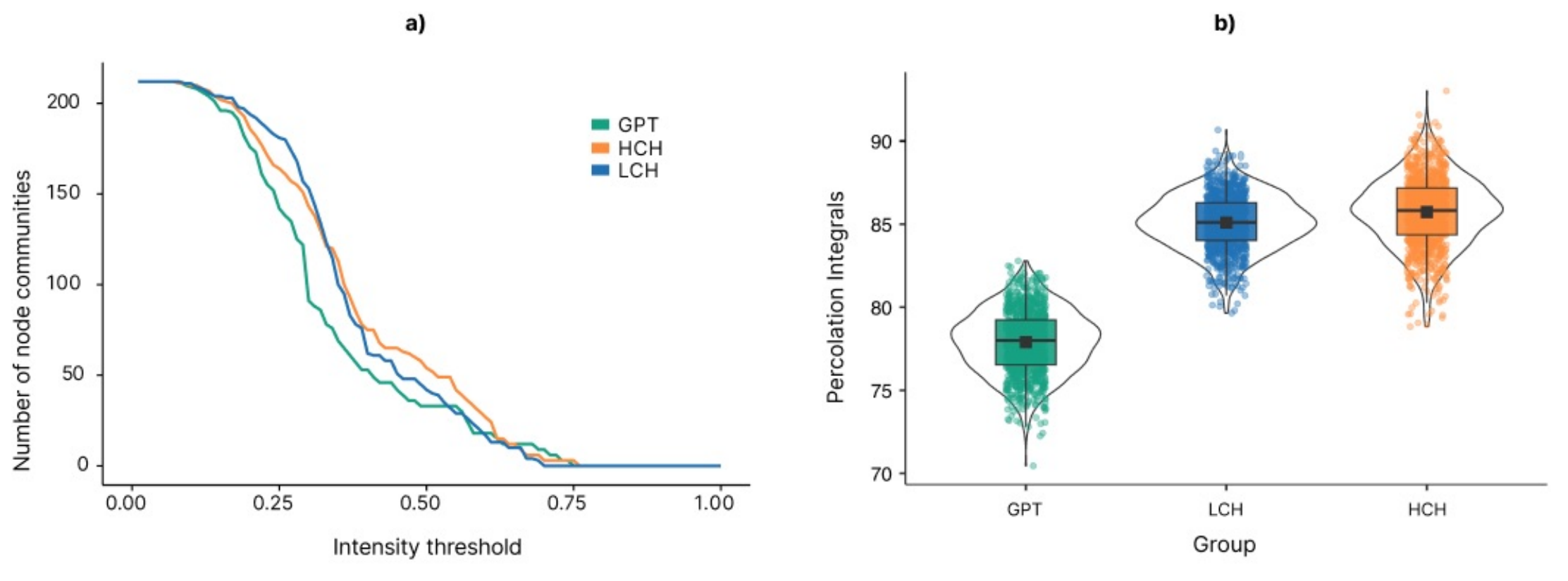}
\caption{Percolation analysis. On the left (a): Representation of the number of connected nodes in each network at increasing thresholds \emph{(I)}. On the right (b): Comparison of the percolation integrals between groups}
  \Description{Figure 5 shows percolation analysis results. There are two figures. The left figure is a graph that shows the number of node communities in ChatGPT-4o, Lower Creative Humans, and Higher Creative Humans networks at increasing intensity thresholds. The X-axis represents the intensity threshold. It ranges from 0.00 to 1.00. The Y-axis shows the number of node communities. It ranges from 0 to 200. Three curves are shown. ChatGPT-4o curve has a lower number of node communities as the intensity threshold increases. Higher Creative Humans curve decreases more gradually, showing a higher number of node communities than ChatGPT-4o as the intensity threshold increases. Lower Creative Humans curve shows an intermediate pattern between ChatGPT-4o and Higher Creative Humans. It has a moderate decrease in the number of node communities as the threshold increases. The right figure represents three violin plots with embedded box plots comparing percolation integrals across the three groups. The Y-axis is percolation integrals, and it ranges from 70 to 90. The X-axis is Group. On the left ChatGPT-4o, in the middle Lower Creative Humans, on the right Higher Creative Humans.}
  \label{fig:percolation}
\end{figure*}
Figure \ref{fig:percolation}a shows how the semantic network of ChatGPT-4o  breaks apart quicker than those of lower and higher creative humans, as it is characterised by a steeper percolation curve and therefore a lower integral \cite{COSGROVE2021104631}. To test for statistical significance, 500 realisations of the percolation test were conducted. Then a One-Way Welch’s ANOVA was performed on the percolation integrals to investigate whether the means of the three groups were significantly different (Figure \ref{fig:percolation}b). Regarding the measure of \emph{Resiliency (R)}, the analysis showed a significant main effect of Group, $F_{(2, 977)} = 2544$, $p < .001$, $\eta_p^2 = 0.84$. Games-Howell post hoc tests revealed that GPT ($M = 77.9$, $SD = 1.98$) showed significantly lower average percolation integrals compared to LCH ($M = 85.2$, $SD = 1.65$), $t_{(967)} = - 63.3$, $p < .001$, $d = 4.01$, and HCH ($M = 85.8$, $SD = 2.08$), $t_{(996)} = - 61.48$, $p < .001$, $d = 3.89$. Moreover, HCH showed significantly higher average integrals than LCH, $t_{(949)} = 5.04$, $p < .001$, $d = 0.32$.

\section {Discussion}\label{sec:discussion}
This study investigated the association between semantic networks and idea originality in ChatGPT-4o and a sample of 81 psychology students divided into higher and lower creative individuals. In the following subsections, we discuss the results in comparison with research on human creativity \cite{benedek_toward_2019, WANG2025100103, kenett_network} and then reflect on the implications of artificial creativity \cite{RUNCO_2023} for CSTs design \cite{Shneiderman, Frich}.

\subsection{Hypotheses testing} 
Extending Artificial Cognition \cite{taylor_artificial_2021} with research on creative cognition \cite{benedek_toward_2019, WANG2025100103, kenett_network}, the study tested the hypothesis of a relationship between 
the semantic network of ChatGPT-4o and its idea originality \emph{(H1)}. To address it, we tested two sub-hypotheses related to the analysed constructs (Table \ref{tab:summary}).

\subsubsection {Main Effects on Idea Originality and Network Flexibility}
Results supported \emph{H1.1} related to idea originality. They showed that ChatGPT-4o was less original than higher creative humans but more original than lower creative humans (Table \ref{tab:summary}). The effect emerged consistently, considering both the total production and the originality peaks. Differently, the results of the structural and percolation analyses only partially confirmed \emph{H1.2} related to network flexibility. Consistent with previous studies \cite{kenett_semantic_2019, kenett_network}, higher creative humans were characterised by a more organised, flexible, and robust semantic memory network compared to lower creative humans. Furthermore, as expected, the network of ChatGPT-4o was found to be less flexible than that of higher creative humans. However, the same pattern also emerged in the comparison with lower creative individuals, who appeared to be more flexible than the machine. 

To summarise, ChatGPT-4o semantic network was characterised by more distances, weaker local interconnectivity and higher modularity compared to those of both lower and higher creative humans. This configuration indicates higher internal rigidity and less cohesion, which reflects the sparser connections between concepts in different sub-communities \cite{WANG2025100103, kenett_network}. It suggests an increased difficulty in ``navigating'' the semantic structures to reach remote concepts and to create original associations between them \cite{beaty_associative_2023, benedek_role_2023}. This conclusion was confirmed by the percolation analysis. It showed that the semantic network of ChatGPT-4o broke apart faster as its intensity thresholds increased. 

\setcounter{table}{5}
\begin{table}
  \caption{Summary of hypotheses and results}
\Description{Table 6 presents a summary of the expected direction of the two hypotheses of this study and the observed results.}
  \label{tab:summary}
  \begin{tabular}{p{2cm} p{2.8cm} p{2.8cm}}
    \toprule
    \textbf{Hypothesis} & \textbf{Expected direction} & \textbf{Result}\\
    \midrule
  Idea \newline Originality & HCH $>$ GPT $>$ LCH & HCH $>$ GPT $>$ LCH\\
     \midrule
   Network \newline Flexibility & HCH $>$ \textbf{GPT $>$ LCH} & HCH $>$ \textbf{LCH $>$ GPT}\\
    \bottomrule
  \end{tabular}
\end{table}

\subsubsection{Association between Idea Originality and Network Flexibility} 
Our main hypothesis \emph{(H1)} stated that the semantic network flexibility provides an indication of idea originality in LLMs. Results revealed that higher creative humans consistently outperformed ChatGPT-4o in terms of originality. This result supports psychological literature suggesting that highly creative individuals, who are characterised by a less spread out, less segregated, more flexible and robust semantic memory network, exceed AI in creative tasks \cite{cropley_creations_2025, koivisto_best_2023}. However, despite a more rigid and segregated network, ChatGPT-4o consistently exceeded lower creative humans in idea originality (Table ~\ref{tab:summary}). 

These results seem to suggest independence between semantic network structures and creative performance in machines. Yet, they may be better understood in light of the Entropy Modulation Theory of Creative Exploration \cite{hills_entropy_2025}. The authors of this theory suggest that even when the semantic memory structure of groups of individuals is equivalent, higher creative ones navigate the network faster by increasing activation variance or lowering production threshold. While in humans this capacity reflects executive control, in ChatGPT-4o it may be mimicked by different hyperparameters of the model, such as temperature, top\_p, max tokens, frequency and presence penalty.

Temperature, for example, defines the level of randomness of text generation, with lower values producing more deterministic responses, and higher values increasing unpredictability \cite{peeperkorn_temperature_2024, cropley_averagecreativity_2025}. However, 
preliminary results suggest a weak effect of temperature on creative outputs, despite it being described as the creativity
parameter in the literature \cite{peeperkorn_temperature_2024}. 
The effect of the other hyperparameters remains largely unexplored and much uncertainty still exists  \cite{cropley_averagecreativity_2025}.
In this study, we tested ChatGPT-4o using its chat interface. Accordingly, although its network was more rigid, ChatGPT-4o may have emerged as more original than lower creative humans because this model configuration allowed the generation of original alternative uses without sacrificing coherence with respect to the AUT prompt.

Another explanation for these inconsistencies could be related to the impact of affective and motivational processes on human creative performance  \cite{agnoli2018, DeRooij, koivisto_best_2023, amabile_dynamic_2016, karwowski_chapter_2017, corazza_davinci, agnoli_emotions_2019}. While we might assume that ChatGPT-4o consistently performed the task at the best of its incredibly fast computational capabilities \cite{koivisto_best_2023}, lower creative humans might have experienced a higher cognitive effort. Furthermore, they were likely more affected by scarce motivation to do the task, low self-efficacy, external pressure of being evaluated, time constraints, or cognitive fatigue than higher creative humans. Consequently, even with a less flexible network, ChatGPT-4o performed the test consistently at its best capacity and generated more original responses than participants with lower creative abilities. Finally, these inconsistent results could be due to the impact of the ChatGPT-4o speed in processing large data structures \cite{koivisto_best_2023}. Even if its network was not flexible, its rapidity in navigating complex information structures might have led to the retrieval of remote associations at a pace that compensated for its structural limitations.

\subsection{Limitations and Future Directions}

This paper provides a first step in linking semantic networks and artificial creativity. The novelty of the approach encompasses a number of limitations, which limit the generalisability of our results. These pertain to the data sample, the study procedure, and the complexity of the creativity construct.

\subsubsection{Data Sample}
A main limitation concerns the type of human sample and the reliance on only ChatGPT-4o. The human participants were psychology students, who may have been more intrinsically motivated to perform the tasks because of their academic background as compared to other individuals. Since motivation is a key dimension of the creative process \cite{DeRooij, agnoli2018, karwowski_chapter_2017, amabile_dynamic_2016, koivisto_best_2023, corazza_davinci}, it is possible that different participants would have led to different results. Furthermore, we only tested ChatGPT-4o using its chat interface \cite{Vinchon}. This choice tested an extreme black-box that improved ecological validity as it represents the typical interface employed by most users. However, the chat interface does not allow to know and modify the values of the model hyperparameters. Future research is needed to compare different models while manipulating hyperparameters, in order to investigate their effects on the correlation pattern between semantic network flexibility and originality. 

\subsubsection{Study Procedure} 
Following a strict application of Artificial Cognition \cite{taylor_artificial_2021} and consistent with \cite{HAASE_2023, grassini_artificial_2025, GUZIK2023100065, cropley_is_2023, Vinchon}, we prompted ChatGPT-4o with the same instructions given to human participants. Yet, research suggests that prompting plays a fundamental role in the potential creativity of GenAI  \cite{Oppenlaender18082025}. There is evidence that LLMs can be incrementally enforced to generate original responses across interactions \cite{goes_2023pushing} and creative outcomes are more likely to occur during roleplay \cite{Grigis_2024, shanahan_role_2023}. Furthermore, both humans and ChatGPT-4o were given the same time constraints (Table \ref{table_appendix}), but we controlled it only for the human sample. Although it has been suggested that ChatGPT may reflect the serial order effect typical of human creativity \cite{beaty_why_2012,stevenson_2022putting, GILHOOLY2024100071}, no rigorous investigation has yet examined how time limitation affects both machine and human performance. Altogether, these considerations highlight the urgency of research on how prompting affects the production of artificial creativity.

\subsubsection{Construct Complexity}
Following previous research, we investigated divergent thinking through the AUT focusing on originality \cite{stevenson_2022putting, bangerl_2024, Bangerl_2025, HAASE_2023, koivisto_best_2023}. AUT is ``regularly used in work on AI creativity'' \cite[p. 2]{Bangerl_2025} (Table \ref{tab:humans_llms_creativity}). Nevertheless, divergent thinking represents only one aspect of the multi-componential construct that is creativity \cite{Corazza_2016, grassini_artificial_2025}, and originality is only one requirement of a creative idea \cite{RUNCO_2023}. Due to its popularity in research, AUT responses may have been included in the training databases of the machine \cite{GILHOOLY2024100071, HAASE_2023}. However, as Haase \& Hanel \cite{HAASE_2023} noted, this possibility does not imply that these words would appear in the generated output. Future research should extend to more ecological methods to assess creativity, including real world activities, and non-verbal, visual, or figurative performance. Additional creativity measures (i.e., flexibility, elaboration) should also be considered to comprehensively investigate both human and artificial creativity. Finally, future studies should clarify the the directionality of the association between network flexibility and creativity.
Altogether, these limitations provide directions for future research on artificial creativity.

\section{Artificial Creativity}\label{sec:implications}
Given all the limitations acknowledged in the previous section, we now reflect on the relevance of artificial creativity \cite{Runco_2025, RUNCO_2023} for CSTs design and evaluation \cite{Frich, Shneiderman}. Following \cite{RUNCO_2023, Runco_2025}, artificial creativity refers to  the potential of a machine to provide output that if generated by humans would be defined as creative. Building on this definition, we provide two operational insights (OPI) and three methodological  recommendations  (MER) for a systematic research agenda on artificial creativity in HCI. 

\begin{itemize}
\item[OPI1]\emph{Artificial Creativity is Average Performance}: The mean originality scores of ChatGPT-4o on the AUT reflected the average human performance \cite{cropley_averagecreativity_2025}, but cut out lower and higher creative individuals. Therefore, the machine outcome was good enough, but it could not achieve excellence.  This result supports and explains previous qualitative research showing how at first encounter artists tended to perceive artificial creativity as inadequate, constrained, or even constraining \cite{Grigis_2024, Cremaschi_2025, Chakrabarty, Ghajargar, deangeli_2025}. In particular, they were concerned about a risk of homogenisation which threatened their professional identity and personal style.  It was feared that GenAI could make creativity less authentic \cite{aru_artificial_2025} and elaborated \cite{Bangerl_2025}, fostering convergence towards similar average results \cite{doshi_generative_nodate, Kumar, Anderson, bruno_three_2025}. On the other hand, this average definition suggests that GenAI could be beneficial, specifically for less creative individuals, whose output becomes more novel, enjoyable, and useful when supported by AI \cite{doshi_generative_nodate}. Therefore, it is of primary importance to consider the user experience and skills when introducing users to artificial creativity. Regardless of their background, all users require new interfaces to bridge human and artificial creativity. Therefore, designers of human-AI systems should consider the average nature of artificial creativity  when designing interfaces connecting human and artificial creativity to ensure that the former  benefits from rather than being hindered by the interaction \cite{Corazza_decalogue, Kumar}.

\item [OPI2]\emph{Artificial Creativity is a Process, not only a Product}: Artificial creativity expands and connects HCI and psychological research by advancing a dynamic definition of artificial creativity \cite{Corazza_2016}. Not only does this definition address the question of whether AI can be creative, but it also attempts to understand the processes embedded in AI mechanisms. The current paper provides preliminary evidence of a partial association between idea originality (as an indication of the product) and semantic networks (as an indication of the process) in ChatGPT-4o.  The results confirmed the expectation derived by human cognition \cite{kenett_network, kenett_semantic_2019} of a relationship between originality and network flexibility only with respect to the comparison with higher creative individuals. Therefore, the potentiality of generating original ideas may not be strictly associated with the semantic structure, but certain hyperparameters may partly compensate for the rigidity of the network and enable the production of original responses. This papers  offers HCI researchers a structured methodology for systematically addressing artificial creativity. Future research should also focus on supporting different stages of the creative process \cite{corazza_cyber-creativity_2025, Runco_2025, ceh2024creativity}, and  how their use affects users' performance in subsequent tasks with and without them \cite{Bangerl_2025, Kumar}.

\item[MER1] \emph{Human Creativity Provides Methodological Scaffolding}: Research on Artificial Cognition draws parallels between the AI black-box and the human mind --- the original black box  \cite{taylor_artificial_2021}. In sharp contrast with the influential Media Equation Theory \cite{Reeves_Media}, artificial creativity does not assume a similarity between machine and human behaviour.  On the contrary, this paper provides evidence of differences in idea originality and its association with semantic memory structure between human and artificial creativity. The evidence enriches related work describing higher level psychological differences in drivers, experiences and norms \cite{RUNCO_2023, Runco_2025, Abraham_2025}. The framework of creative cognition \cite{benedek_toward_2019} provides concepts, methods, and theories to approach the study of artificial creativity. From a design perspective, it addresses the interface at a micro-level, directly connecting machine hyperparameters and human executive functions. In this paper we have analysed semantic memory, future research should address attention and control (Figure ~\ref{fig:Creative Cognition}), as relevant dimensions of both human and artificial creativity.

\item [MER2]\emph{Collecting Reliable Human Baselines}: The human sample used as a baseline is of critical importance for the study of artificial creativity. From a cognitive perspective, indeed, creativity emerges as the expression of personal abilities, which vary across individuals and along the lifespan \cite{RUNCO_2023}. Furthermore, the natural black box is affected by hidden cognitive functions, motivations, and experiences, all of which can affect creative performance. Consequently, an average representation of creativity only applies to the machine, but it fails to explain the richness of the human counterpart (Figures \ref{fig:teaser} and \ref{fig:density}). It makes limited sense to talk about the average creativity of humankind. Consequently, it needs to be better characterised taking into account possibilities and limitations. The median split applied in this study offers a simple way to obtain a finer-grained operationalisation of creativity for future research.  Other individual and contextual factors also need careful consideration. They relate to participants' recruitment and experimental design.  

\item [MER3]\emph{Artificial Creativity Provides Metaphors for Design}: Over 20 years ago, Shneiderman \cite{Shneiderman} suggested that ``Graphics composition tools show the enormous power of software to enable more people to be more creative '' \cite[p. 119]{Shneiderman}.  Consistently, the metaphor of semantic network opens up stimulating design trajectories for HCI research.  We can imagine new CSTs allowing different types of users (artists and researchers) to visualise and manipulate the semantic networks of the model. By modifying these networks as creative interfaces, artists may explore new semantic spaces which may lead to seed incidents at the core of their masterpieces \cite{deangeli_2025}. There is evidence indeed that experimental artists engage in experiments with extensive data curation \cite{caramiaux}, subverting conceptual spaces, forcing the machine to fail and hallucinate \cite{Bomba_2024, Bomba_2025}, or introducing deliberate frictions in tangible interactions \cite{ Cremaschi_steamp}. Experimental artists may value the tool as a medium for transformational practices \cite{Bomba_2025, caramiaux, Bomba_2024}. Scientists, on the other hand, may exploit the interactive visualisation of semantic networks to support a systematic program of research on artificial creativity. They could manipulate prompts, hyperparameters, and tasks to understand their effect on the system output.  This research requires different methods and sensibilities when testing humans or machines.  Current research has paid extensive attention to the prompt provided to the machine \cite{Oppenlaender18082025, goes_2023pushing}. This paper instead highlighted methodological considerations for user research.

\end{itemize}

\section{Conclusion}\label{sec:conclusion}
To conclude, we go back to the question posed in the introduction about the {\emph{the mechanisms that could lead to artificial creativity.} Extending the framework of Artificial Cognition \cite{taylor_artificial_2021} to account for creativity \cite{benedek_toward_2019}, this paper provides initial evidence of the role of semantic networks on idea originality in LLMs using the the Alternate Uses Task. We demonstrated that this role differed according to the human sample used for the comparison. Higher creative individuals were more original than ChatGPT-4o, which, despite its rigidity, emerged as more original than lower creative individuals. 
Taken together, the evidence extends previous studies comparing human and artificial creativity by providing operational and methodological advances. Finally, we provide methodological guidance for a more rigorous investigation of artificial creativity in the field of HCI. }

\begin{acks}
The authors are grateful to all participants who took part in the study. We thank Fe Simeoni and Moritz Mock for their availability and for discussions of technical aspects. The second author is supported by a scholarship funded by the European Union – NextGenerationEU and the Italian Ministry of University and Research (MUR), CUP J92B23001680007. We acknowledge financial support by European Union – NextGenerationEU, CUP J53D2300799 0001.
\end{acks}

\bibliographystyle{ACM-Reference-Format}
\bibliography{sample-base}

@String{Computing = "Computing" }

@String{Computer = "{IEEE} Computer" }

@String{Academic = "Academic Press" }

@String{Springer = "Springer-Verlag" }

@article{Abraham_2025,
author = {Anna Abraham},
title ={Why the standard definition of creativity fails to capture the creative act},
journal = {Theory \& Psychology},
volume = {35},
number = {1},
pages = {40-60},
year = {2025},
doi = {10.1177/09593543241290232},
}

@article{agnoli_dopamine_2023,
	title = {Dopamine supports idea originality: the role of spontaneous eye blink rate on divergent thinking},
	volume = {87},
	issn = {1430-2772},
	url = {https://doi.org/10.1007/s00426-022-01658-y},
	doi = {10.1007/s00426-022-01658-y},
	number = {1},
	journal = {Psychological Research},
	author = {Agnoli, Sergio and Mastria, Serena and Zanon, Marco and Corazza, Giovanni Emanuele},
	month = feb,
	year = {2023},
	pages = {17--27},
}

@article{Agnoli_2022,
author = {Agnoli, Sergio and Pozzoli, Tiziana and Mancini, Giacomo and Franchin, Laura and Mastria, Serena and Corazza, Giovanni Emanuele},
title = {This is My Fairy Tale: How Emotional Intelligence Interacts with a Training Intervention in Enhancing Children's Creative Potential},
journal = {The Journal of Creative Behavior},
volume = {56},
number = {3},
pages = {465-482},
doi = {https://doi.org/10.1002/jocb.541},
url = {https://onlinelibrary.wiley.com/doi/abs/10.1002/jocb.541},
year = {2022}
}

@inproceedings{Bangerl_2025,
author = {Bangerl, Mia Magdalena and Disch, Leonie and David, Tamara and Pammer-Schindler, Viktoria},
title = {CreAItive Collaboration? Users' Misjudgment of AI-Creativity Affects Their Collaborative Performance},
year = {2025},
isbn = {9798400713941},
publisher = {Association for Computing Machinery},
address = {New York, NY, USA},
url = {https://doi.org/10.1145/3706598.3713886},
doi = {10.1145/3706598.3713886},
booktitle = {Proceedings of the 2025 CHI Conference on Human Factors in Computing Systems},
articleno = {195},
numpages = {17},
location = {
},
series = {CHI '25}
}

@article{bangerl_2024,
  title={Explorations in human vs. generative AI creative performances: A study on human-AI creative potential},
  author={Bangerl, Mia Magdalena and Stefan, Katharina and Pammer-Schindler, Viktoria},
  journal={TREW 2024: Trust and Reliance in Evolving Human-AI Workflows, at CHI 2024, May 11},
  year={2024}
}

@article{beaty_why_2012,
	title = {Why {Do} {Ideas} {Get} {More} {Creative} {Across} {Time}? {An} {Executive} {Interpretation} of the {Serial} {Order} {Effect} in {Divergent} {Thinking} {Tasks}},
	volume = {6},
	shorttitle = {Why {Do} {Ideas} {Get} {More} {Creative} {Across} {Time}?},
	doi = {10.1037/a0029171},
	journal = {Psychology of Aesthetics Creativity and the Arts},
	author = {Beaty, Roger and Silvia, Paul},
	month = aug,
	year = {2012},
	pages = {309--319},
}

@article{beaty_associative_2023,
	title = {Associative thinking at the core of creativity},
	volume = {27},
	issn = {1364-6613},
	url = {https://doi.org/10.1016/j.tics.2023.04.004},
	doi = {10.1016/j.tics.2023.04.004},
	number = {7},
	urldate = {2025-08-28},
	journal = {Trends in Cognitive Sciences},
	author = {Beaty, Roger E. and Kenett, Yoed N.},
	month = jul,
	year = {2023},
	note = {Publisher: Elsevier},
	pages = {671--683},
}

@article{benedek_role_2023,
	title = {The role of memory in creative ideation},
	volume = {2},
	issn = {2731-0574},
	url = {https://doi.org/10.1038/s44159-023-00158-z},
	doi = {10.1038/s44159-023-00158-z},
	number = {4},
	journal = {Nature Reviews Psychology},
	author = {Benedek, Mathias and Beaty, Roger E. and Schacter, Daniel L. and Kenett, Yoed N.},
	month = apr,
	year = {2023},
	pages = {246--257},
}

@article{benedek_toward_2019,
	title = {Toward a neurocognitive framework of creative cognition: the role of memory, attention, and cognitive control},
	volume = {27},
	issn = {2352-1546},
	url = {https://www.sciencedirect.com/science/article/pii/S2352154618301839},
	doi = {https://doi.org/10.1016/j.cobeha.2018.11.002},
	journal = {Current Opinion in Behavioral Sciences},
	author = {Benedek, Mathias and Fink, Andreas},
	year = {2019},
	pages = {116--122},
}

@article{Bomba_2024,
author = {Bomba, Federico and Men\'{e}ndez-Blanco, Mar\'{\i}a and Grigis, Paolo and Cremaschi, Michele and De Angeli, Antonella},
title = {The Choreographer-Performer Continuum: A Diffraction Tool to Illuminate Authorship in More Than Human Co-Performances},
year = {2024},
issue_date = {December 2024},
publisher = {Association for Computing Machinery},
address = {New York, NY, USA},
volume = {31},
number = {6},
issn = {1073-0516},
url = {https://doi.org/10.1145/3689040},
doi = {10.1145/3689040},
journal = {ACM Trans. Comput.-Hum. Interact.},
month = dec,
articleno = {75},
numpages = {23}
}

@article{BORODKIN,
title = {When pumpkin is closer to onion than to squash: The structure of the second language lexicon},
journal = {Cognition},
volume = {156},
pages = {60-70},
year = {2016},
issn = {0010-0277},
doi = {https://doi.org/10.1016/j.cognition.2016.07.014},
url = {https://www.sciencedirect.com/science/article/pii/S0010027716301895},
author = {Katy Borodkin and Yoed N. Kenett and Miriam Faust and Nira Mashal}
}

@article{farkas_weighted_2007,
	title = {Weighted network modules},
	volume = {9},
	issn = {1367-2630},
	url = {https://dx.doi.org/10.1088/1367-2630/9/6/180},
	doi = {10.1088/1367-2630/9/6/180},
	number = {6},
	journal = {New Journal of Physics},
	author = {Farkas, Illés and Ábel, Dániel and Palla, Gergely and Vicsek, Tamás},
	month = jun,
	year = {2007},
	pages = {180},
}

@article{palla_uncovering_2005,
	title = {Uncovering the overlapping community structure of complex networks in nature and society},
	volume = {435},
	issn = {1476-4687},
	url = {https://doi.org/10.1038/nature03607},
	doi = {10.1038/nature03607},
	number = {7043},
	journal = {Nature},
	author = {Palla, Gergely and Derényi, Imre and Farkas, Illés and Vicsek, Tamás},
	month = jun,
	year = {2005},
	pages = {814--818},
}

@article{Shao,
author = {Jun Shao},
title = {{Impact of the Bootstrap on Sample Surveys}},
volume = {18},
journal = {Statistical Science},
number = {2},
publisher = {Institute of Mathematical Statistics},
pages = {191 -- 198},
year = {2003},
doi = {10.1214/ss/1063994974},
URL = {https://doi.org/10.1214/ss/1063994974}
}

@article{politis_large_1994,
	title = {Large {Sample} {Confidence} {Regions} {Based} on {Subsamples} under {Minimal} {Assumptions}},
	volume = {22},
	issn = {0090-5364, 2168-8966},
	url = {https://projecteuclid.org/journals/annals-of-statistics/volume-22/issue-4/Large-Sample-Confidence-Regions-Based-on-Subsamples-under-Minimal-Assumptions/10.1214/aos/1176325770.full},
	doi = {10.1214/aos/1176325770},
	number = {4},
	urldate = {2025-08-18},
	journal = {The Annals of Statistics},
	author = {Politis, Dimitris N. and Romano, Joseph P.},
	month = dec,
	year = {1994},
	note = {Publisher: Institute of Mathematical Statistics},
	pages = {2031--2050},
}

@article{doshi_generative_nodate,
	title = {Generative {AI} enhances individual creativity but reduces the collective diversity of novel content},
	volume = {10},
	url = {https://doi.org/10.1126/sciadv.adn5290},
	doi = {10.1126/sciadv.adn5290},
	number = {28},
year={2024},
	urldate = {2025-08-20},
	journal = {Science Advances},
	author = {Doshi, Anil R. and Hauser, Oliver P.},
	note = {Publisher: American Association for the Advancement of Science},
	pages = {eadn5290},
}

@article{caramiaux,
author = {Caramiaux, Baptiste and Fdili Alaoui, Sarah},
title = {"Explorers of Unknown Planets": Practices and Politics of Artificial Intelligence in Visual Arts},
year = {2022},
issue_date = {November 2022},
publisher = {Association for Computing Machinery},
address = {New York, NY, USA},
volume = {6},
number = {CSCW2},
url = {https://doi.org/10.1145/3555578},
doi = {10.1145/3555578},
journal = {Proc. ACM Hum.-Comput. Interact.},
month = nov,
articleno = {477},
numpages = {24},
}

@book{Reeves_Media,
author = {Reeves, Byron and Nass, Clifford},
title = {The media equation: how people treat computers, television, and new media like real people and places},
year = {1996},
isbn = {157586052X},
publisher = {Cambridge University Press},
address = {USA}
}

@article{sternberg_investing_1996,
	title = {Investing in creativity},
	volume = {51},
	issn = {1935-990X},
	doi = {10.1037/0003-066X.51.7.677},
	number = {7},
	journal = {American Psychologist},
	author = {Sternberg, Robert J. and Lubart, Todd I.},
	year = {1996},
Publisher = {American Psychological Association},
	pages = {677--688},
}

@article{Bomba_2025,
	title = {Agency and authorship in {AI} art: {Transformational} practices for epistemic troubles},
	volume = {205},
	issn = {1071-5819},
	url = {https://www.sciencedirect.com/science/article/pii/S1071581925002095},
	doi = {https://doi.org/10.1016/j.ijhcs.2025.103652},
	journal = {International Journal of Human-Computer Studies},
	author = {Bomba, Federico and Angeli, Antonella De},
	year = {2025},
	pages = {103652},
}

@article{bertail_second-order_1997,
	title = {Second-{Order} {Properties} of an {Extrapolated} {Bootstrap} without {Replacement} under {Weak} {Assumptions}},
	volume = {3},
	issn = {13507265},
	url = {http://www.jstor.org/stable/3318585},
	doi = {10.2307/3318585},
	number = {2},
	urldate = {2025-08-20},
	journal = {Bernoulli},
	author = {Bertail, Patrice},
	year = {1997},
	note = {Publisher: International Statistical Institute (ISI) and Bernoulli Society for Mathematical Statistics and Probability},
	pages = {149--179},
}

@article{ceh2024creativity,
  title={Creativity in digitally mediated times: How digital tools support creativity across domains.},
  author={Ceh, Simon M and Rafner, Janet and Benedek, Mathias},
  journal={Psychology of Aesthetics, Creativity, and the Arts},
  year={2024},
  publisher={Educational Publishing Foundation}
}

@article{christensen_tutorial2023,
	title = {Semantic network analysis ({SemNA}): {A} tutorial on preprocessing, estimating, and analyzing semantic networks},
	volume = {28},
	issn = {1939-1463},
	shorttitle = {Semantic network analysis ({SemNA})},
	doi = {10.1037/met0000463},
	number = {4},
	journal = {Psychological Methods},
	author = {Christensen, Alexander P. and Kenett, Yoed N.},
	year = {2023},
	note = {Place: US
Publisher: American Psychological Association},
	pages = {860--879},
}

@article{b_efron_bootstrap_1979,
	title = {Bootstrap {Methods}: {Another} {Look} at the {Jackknife}},
	volume = {7},
	url = {https://doi.org/10.1214/aos/1176344552},
	doi = {10.1214/aos/1176344552},
	number = {1},
	journal = {The Annals of Statistics},
	author = {{B. Efron}},
	month = jan,
	year = {1979},
	pages = {1--26},
}

@article{agnoli2018,
	title = {The role of motivation in the prediction of creative achievement inside and outside of school environment},
	volume = {28},
	issn = {1871-1871},
	url = {https://www.sciencedirect.com/science/article/pii/S1871187117303528},
	doi = {https://doi.org/10.1016/j.tsc.2018.05.005},
	journal = {Thinking Skills and Creativity},
	author = {Agnoli, Sergio and Runco, Mark A. and Kirsch, Christiane and Corazza, Giovanni Emanuele},
	year = {2018},
	pages = {167--176},
}

@article{rastelli_hallucinations,
	title = {Simulated visual hallucinations in virtual reality enhance cognitive flexibility},
	volume = {12},
	issn = {2045-2322},
	url = {https://doi.org/10.1038/s41598-022-08047-w},
	doi = {10.1038/s41598-022-08047-w},
	number = {1},
	journal = {Scientific Reports},
	author = {Rastelli, Clara and Greco, Antonino and Kenett, Yoed N. and Finocchiaro, Chiara and De Pisapia, Nicola},
	month = mar,
	year = {2022},
	pages = {4027},
}

@article{lubart2013creative,
  title={Creative potential and its measurement.},
  author={Lubart, Todd and Zenasni, Franck and Barbot, Baptiste},
  journal={International Journal for Talent Development and Creativity},
  volume={1},
  number={2},
  pages={41--51},
  year={2013},
  publisher={ERIC}
}

@incollection{corazza_davinci,
	address = {Cham},
	title = {The {DA} {VINCI} {Model} for the {Creative} {Thinking} {Process}},
	isbn = {978-3-030-99674-1},
	url = {https://doi.org/10.1007/978-3-030-99674-1_4},
	booktitle = {Homo {Creativus}: {The} 7 {C}'s of {Human} {Creativity}},
	publisher = {Springer International Publishing},
	author = {Corazza, Giovanni Emanuele and Agnoli, Sergio},
	editor = {Lubart, Todd and Botella, Marion and Bourgeois -Bougrine, Samira and Caroff, Xavier and Guegan, Jerome and Mouchiroud, Christophe and Nelson, Julien and Zenasni, Franck},
	year = {2022},
	doi = {10.1007/978-3-030-99674-1_4},
	pages = {49--67},
}

@article{gabora2018neural,
  title={The neural basis and evolution of divergent and convergent thought},
  author={Gabora, Liane},
  journal={The Cambridge handbook of the neuroscience of creativity},
  pages={58--70},
  year={2018},
  publisher={Cambridge University Press Cambridge}
}

@article{christensen_network_2018,
	title = {Network structure of the {Wisconsin} {Schizotypy} {Scales}–{Short} {Forms}: {Examining} psychometric network filtering approaches},
	volume = {50},
	issn = {1554-3528},
	url = {https://doi.org/10.3758/s13428-018-1032-9},
	doi = {10.3758/s13428-018-1032-9},
	number = {6},
	journal = {Behavior Research Methods},
	author = {Christensen, Alexander P. and Kenett, Yoed N. and Aste, Tomaso and Silvia, Paul J. and Kwapil, Thomas R.},
	month = dec,
	year = {2018},
	pages = {2531--2550},
}

@article{Corazza_2016,
author = {Giovanni Emanuele Corazza},
title = {Potential Originality and Effectiveness: The Dynamic Definition of Creativity},
journal = {Creativity Research Journal},
volume = {28},
number = {3},
pages = {258--267},
year = {2016},
publisher = {Routledge},
doi = {10.1080/10400419.2016.1195627},
}

@inproceedings{Kumar,
author = {Kumar, Harsh and Vincentius, Jonathan and Jordan, Ewan and Anderson, Ashton},
title = {Human Creativity in the Age of LLMs: Randomized Experiments on Divergent and Convergent Thinking},
year = {2025},
isbn = {9798400713941},
publisher = {Association for Computing Machinery},
address = {New York, NY, USA},
url = {https://doi.org/10.1145/3706598.3714198},
doi = {10.1145/3706598.3714198},
booktitle = {Proceedings of the 2025 CHI Conference on Human Factors in Computing Systems},
articleno = {23},
numpages = {18},
location = {
},
series = {CHI '25}
}

@incollection{bruno_three_2025,
	address = {Cham},
	title = {The {Three} {Waves} {Paradigm}: {Tracing} the {Evolution} of {Artificial} {Intelligence} in {Design}},
	isbn = {978-3-031-84535-2},
	url = {https://doi.org/10.1007/978-3-031-84535-2_5},
	booktitle = {The {Cyber}-{Creativity} {Process}: {How} {Humans} {Co}-{Create} with {Artificial} {Intelligence}},
	publisher = {Springer Nature Switzerland},
	author = {Bruno, Carmen and Canina, Marita and Biskjaer, Michael Mose},
	editor = {Corazza, Giovanni Emanuele},
	year = {2025},
	doi = {10.1007/978-3-031-84535-2_5},
	pages = {129--162},
}

@inproceedings{ning_CST_2023,
	title = {Creativity {Support} in {AI} {Co}-creative {Tools}: {Current} {Research}, {Challenges} and {Opportunities}},
	shorttitle = {Creativity {Support} in {AI} {Co}-creative {Tools}},
	url = {https://ieeexplore.ieee.org/document/10152832/},
	doi = {10.1109/CSCWD57460.2023.10152832},
	urldate = {2025-11-28},
	booktitle = {2023 26th {International} {Conference} on {Computer} {Supported} {Cooperative} {Work} in {Design} ({CSCWD})},
	author = {Ning, Bin and Liu, Fang and Liu, Zhixiong},
	month = may,
	year = {2023},
	note = {ISSN: 2768-1904},
	pages = {5--10},
}

@inproceedings{chung_AICST_DIS2021,
	address = {Virtual Event USA},
	title = {The {Intersection} of {Users}, {Roles}, {Interactions}, and {Technologies} in {Creativity} {Support} {Tools}},
	isbn = {978-1-4503-8476-6},
	url = {https://dl.acm.org/doi/10.1145/3461778.3462050},
	doi = {10.1145/3461778.3462050},
	language = {en},
	urldate = {2025-11-28},
	booktitle = {Designing {Interactive} {Systems} {Conference} 2021},
	publisher = {ACM},
	author = {Chung, John Joon Young and He, Shiqing and Adar, Eytan},
	month = jun,
	year = {2021},
	pages = {1817--1833},
}

@incollection{agnoli_emotions_2019,
	address = {Cham},
	title = {Emotions: {The} {Spinal} {Cord} of the {Creative} {Thinking} {Process}},
	isbn = {978-3-319-99163-4},
	url = {https://doi.org/10.1007/978-3-319-99163-4_4},
	booktitle = {Dynamic {Perspectives} on {Creativity}: {New} {Directions} for {Theory}, {Research}, and {Practice} in {Education}},
	publisher = {Springer International Publishing},
	author = {Agnoli, Sergio and Corazza, Giovanni Emanuele},
	editor = {Beghetto, Ronald A. and Corazza, Giovanni Emanuele},
	year = {2019},
	doi = {10.1007/978-3-319-99163-4_4},
	pages = {47--65},
}

@book{corazza_cyber-creativity_2025,
	address = {Cham},
	title = {The {Cyber}-{Creativity} {Process}: {How} {Humans} {Co}-{Create} with {Artificial} {Intelligence}},
	copyright = {https://www.springernature.com/gp/researchers/text-and-data-mining},
	isbn = {978-3-031-84534-5 978-3-031-84535-2},
	shorttitle = {The {Cyber}-{Creativity} {Process}},
	url = {https://link.springer.com/10.1007/978-3-031-84535-2},
	language = {en},
	urldate = {2025-04-28},
	publisher = {Springer Nature Switzerland},
	editor = {Corazza, Giovanni Emanuele},
	year = {2025},
	doi = {10.1007/978-3-031-84535-2},
	
}

@Article{Corazza_decalogue,
AUTHOR = {Corazza, Giovanni Emanuele and Agnoli, Sergio and Jorge Artigau, Ana and Beghetto, Ronald A. and Bonnardel, Nathalie and Coletto, Irene and Faiella, Angela and Gerardini, Katusha and Gilhooly, Kenneth and Glăveanu, Vlad P. and Hanson, Michael Hanchett and Kapoor, Hansika and Kaufman, James C. and Kenett, Yoed N. and Kharkhurin, Anatoliy V. and Luchini, Simone and Mangion, Margaret and Mirabile, Mario and Obialo, Felix-Kingsley and Phelps, Connie and Reiter-Palmon, Roni and Puryear, Jeb S. and Sarcinella, Eleonora Diletta and Tang, Min and Vavassori, Giulia Maria and Vinchon, Florent and Viskontas, Indre and Weiss, Selina and Zbainos, Dimitrios and Lubart, Todd},
TITLE = {Cyber-Creativity: A Decalogue of Research Challenges},
JOURNAL = {Journal of Intelligence},
VOLUME = {13},
YEAR = {2025},
NUMBER = {8},
ARTICLE-NUMBER = {103},
URL = {https://www.mdpi.com/2079-3200/13/8/103},
ISSN = {2079-3200},
DOI = {10.3390/jintelligence13080103}
}

@article{Corazza_2022,
author = {Corazza, Giovanni Emanuele and Agnoli, Sergio and Mastria, Serena},
title = {The Dynamic Creativity Framework},
journal = {European Psychologist},
volume = {27},
number = {3},
pages = {191-206},
year = {2022},
doi = {10.1027/1016-9040/a000473}
}

@article{COSGROVE2021104631,
title = {Quantifying flexibility in thought: The resiliency of semantic networks differs across the lifespan},
journal = {Cognition},
volume = {211},
pages = {104631},
year = {2021},
issn = {0010-0277},
doi = {https://doi.org/10.1016/j.cognition.2021.104631},
url = {https://www.sciencedirect.com/science/article/pii/S0010027721000500},
author = {Abigail L. Cosgrove and Yoed N. Kenett and Roger E. Beaty and Michele T. Diaz}
}

@article{cropley_is_2023,
	title = {Is artificial intelligence more creative than humans?: {ChatGPT} and the {Divergent} {Association} {Task}},
	volume = {2},
	issn = {2981-877X},
	shorttitle = {Is artificial intelligence more creative than humans?},
	url = {https://learningletters.org/index.php/learn/article/view/13},
	doi = {10.59453/ll.v2.13},
	language = {en},
	urldate = {2025-01-08},
	journal = {Learning Letters},
	author = {Cropley, David},
	month = sep,
	year = {2023},
	pages = {13},
}

@book{cropley_creations_2025,
	address = {Cham},
	title = {Creations: {The} {Nature} of {Creative} {Products} in the 21st {Century}},
	copyright = {https://www.springernature.com/gp/researchers/text-and-data-mining},
	isbn = {978-3-031-82414-2 978-3-031-82415-9},
	shorttitle = {Creations},
	url = {https://link.springer.com/10.1007/978-3-031-82415-9},
	language = {en},
	urldate = {2025-03-28},
	publisher = {Springer Nature Switzerland},
	editor = {Cropley, David H},
	year = {2025},
	doi = {10.1007/978-3-031-82415-9},
}

@article{DeRooij,
author = {de Rooij, Alwin and Vromans, Ruben D.},
title = {The (Dis)Pleasures of Creativity: Spontaneous Eye Blink Rate during Divergent and Convergent Thinking Depends on Individual Differences in Positive and Negative Affect},
journal = {The Journal of Creative Behavior},
volume = {54},
number = {2},
pages = {436-452},
doi = {https://doi.org/10.1002/jocb.379},
url = {https://onlinelibrary.wiley.com/doi/abs/10.1002/jocb.379},
year = {2020},
}

@inproceedings{Frich_CSTsHCI,
author = {Frich, Jonas and Mose Biskjaer, Michael and Dalsgaard, Peter},
title = {Twenty Years of Creativity Research in Human-Computer Interaction: Current State and Future Directions},
year = {2018},
isbn = {9781450351980},
publisher = {Association for Computing Machinery},
address = {New York, NY, USA},
url = {https://doi.org/10.1145/3196709.3196732},
doi = {10.1145/3196709.3196732},
booktitle = {Proceedings of the 2018 Designing Interactive Systems Conference},
pages = {1235–1257},
numpages = {23},
location = {Hong Kong, China},
series = {DIS '18}
}

@article{fischer_CST,
	title = {Creativity {Support} {Tools}: {Report} {From} a {U}.{S}. {National} {Science} {Foundation} {Sponsored} {Workshop}},
	volume = {20},
	issn = {1044-7318, 1532-7590},
	shorttitle = {Creativity {Support} {Tools}},
	url = {http://www.tandfonline.com/doi/abs/10.1207/s15327590ijhc2002_1},
	doi = {10.1207/s15327590ijhc2002_1},
	language = {en},
	number = {2},
	urldate = {2025-11-28},
	journal = {International Journal of Human-Computer Interaction},
	author = {Shneiderman, Ben and Fischer, Gerhard and Czerwinski, Mary and Resnick, Mitch and Myers, Brad and Candy, Linda and Edmonds, Ernest and Eisenberg, Mike and Giaccardi, Elisa and Hewett, Tom and Jennings, Pamela and Kules, Bill and Nakakoji, Kumiyo and Nunamaker, Jay and Pausch, Randy and Selker, Ted and Sylvan, Elisabeth and Terry, Michael},
	month = may,
	year = {2006},
	pages = {61--77},
}

@inproceedings{Frich,
author = {Frich, Jonas and MacDonald Vermeulen, Lindsay and Remy, Christian and Biskjaer, Michael Mose and Dalsgaard, Peter},
title = {Mapping the Landscape of Creativity Support Tools in HCI},
year = {2019},
isbn = {9781450359702},
publisher = {Association for Computing Machinery},
address = {New York, NY, USA},
url = {https://doi.org/10.1145/3290605.3300619},
doi = {10.1145/3290605.3300619},
booktitle = {Proceedings of the 2019 CHI Conference on Human Factors in Computing Systems},
pages = {1–18},
numpages = {18},
location = {Glasgow, Scotland Uk},
series = {CHI '19}
}

@book{geroimenko_human-computer_2025,
	address = {Cham},
	series = {Springer {Series} on {Cultural} {Computing}},
	title = {Human-{Computer} {Creativity}: {Generative} {AI} in {Education}, {Art}, and {Healthcare}},
	copyright = {https://www.springernature.com/gp/researchers/text-and-data-mining},
	isbn = {978-3-031-86550-3 978-3-031-86551-0},
	shorttitle = {Human-{Computer} {Creativity}},
	url = {https://link.springer.com/10.1007/978-3-031-86551-0},
	language = {en},
	urldate = {2025-06-19},
	publisher = {Springer Nature Switzerland},
	editor = {Geroimenko, Vladimir},
	year = {2025},
	doi = {10.1007/978-3-031-86551-0},
}

@article{GILHOOLY2024100071,
title = {AI vs humans in the AUT: Simulations to LLMs},
journal = {Journal of Creativity},
volume = {34},
number = {1},
pages = {100071},
year = {2024},
issn = {2713-3745},
doi = {https://doi.org/10.1016/j.yjoc.2023.100071},
url = {https://www.sciencedirect.com/science/article/pii/S2713374523000304},
author = {Ken Gilhooly}
}

@article{goni_semantic_2011,
	title = {The semantic organization of the animal category: evidence from semantic verbal fluency and network theory},
	volume = {12},
	copyright = {http://www.springer.com/tdm},
	issn = {1612-4782, 1612-4790},
	shorttitle = {The semantic organization of the animal category},
	url = {http://link.springer.com/10.1007/s10339-010-0372-x},
	doi = {10.1007/s10339-010-0372-x},
	language = {en},
	number = {2},
	urldate = {2025-07-14},
	journal = {Cognitive Processing},
	author = {Goñi, Joaquín and Arrondo, Gonzalo and Sepulcre, Jorge and Martincorena, Iñigo and Vélez De Mendizábal, Nieves and Corominas-Murtra, Bernat and Bejarano, Bartolomé and Ardanza-Trevijano, Sergio and Peraita, Herminia and Wall, Dennis P. and Villoslada, Pablo},
	month = may,
	year = {2011},
	note = {Publisher: Springer Science and Business Media LLC},
	pages = {183--196},
}

@inproceedings{Grigis_2024,
author = {Grigis, Paolo and De Angeli, Antonella},
title = {Playwriting with Large Language Models: Perceived Features, Interaction Strategies and Outcomes},
year = {2024},
isbn = {9798400717642},
publisher = {Association for Computing Machinery},
address = {New York, NY, USA},
url = {https://doi.org/10.1145/3656650.3656688},
doi = {10.1145/3656650.3656688},
booktitle = {Proceedings of the 2024 International Conference on Advanced Visual Interfaces},
articleno = {38},
numpages = {9},
location = {Arenzano, Genoa, Italy},
series = {AVI '24}
}

@article{bates_fitting_2015,
	title = {Fitting {Linear} {Mixed}-{Effects} {Models} {Using} lme4},
	volume = {67},
	url = {https://www.jstatsoft.org/index.php/jss/article/view/v067i01},
	doi = {10.18637/jss.v067.i01},
	number = {1},
	journal = {Journal of Statistical Software},
	author = {Bates, Douglas and Mächler, Martin and Bolker, Ben and Walker, Steve},
	year = {2015},
	pages = {1--48},
}

@inproceedings{Remy_CSTsHCI,
author = {Remy, Christian and MacDonald Vermeulen, Lindsay and Frich, Jonas and Biskjaer, Michael Mose and Dalsgaard, Peter},
title = {Evaluating Creativity Support Tools in HCI Research},
year = {2020},
isbn = {9781450369749},
publisher = {Association for Computing Machinery},
address = {New York, NY, USA},
url = {https://doi.org/10.1145/3357236.3395474},
doi = {10.1145/3357236.3395474},
booktitle = {Proceedings of the 2020 ACM Designing Interactive Systems Conference},
pages = {457–476},
numpages = {20},
location = {Eindhoven, Netherlands},
series = {DIS '20}
}

@inproceedings{brown_language_2020,
	title = {Language {Models} are {Few}-{Shot} {Learners}},
	volume = {33},
	url = {https://proceedings.neurips.cc/paper_files/paper/2020/file/1457c0d6bfcb4967418bfb8ac142f64a-Paper.pdf},
	booktitle = {Advances in {Neural} {Information} {Processing} {Systems}},
	publisher = {Curran Associates, Inc.},
	author = {Brown, Tom and Mann, Benjamin and Ryder, Nick and Subbiah, Melanie and Kaplan, Jared D and Dhariwal, Prafulla and Neelakantan, Arvind and Shyam, Pranav and Sastry, Girish and Askell, Amanda and Agarwal, Sandhini and Herbert-Voss, Ariel and Krueger, Gretchen and Henighan, Tom and Child, Rewon and Ramesh, Aditya and Ziegler, Daniel and Wu, Jeffrey and Winter, Clemens and Hesse, Chris and Chen, Mark and Sigler, Eric and Litwin, Mateusz and Gray, Scott and Chess, Benjamin and Clark, Jack and Berner, Christopher and McCandlish, Sam and Radford, Alec and Sutskever, Ilya and Amodei, Dario},
	editor = {Larochelle, H. and Ranzato, M. and Hadsell, R. and Balcan, M. F. and Lin, H.},
	year = {2020},
	pages = {1877--1901},
}

@article{kuznetsova_lmertest_2017,
	title = {{lmerTest} {Package}: {Tests} in {Linear} {Mixed} {Effects} {Models}},
	volume = {82},
	url = {https://www.jstatsoft.org/index.php/jss/article/view/v082i13},
	doi = {10.18637/jss.v082.i13},
	number = {13},
	journal = {Journal of Statistical Software},
	author = {Kuznetsova, Alexandra and Brockhoff, Per B. and Christensen, Rune H. B.},
	year = {2017},
	pages = {1--26},
}

@article{GUILFORD,
author = {GUILFORD, J. P.},
title = {Creativity: Yesterday, Today and Tomorrow},
journal = {The Journal of Creative Behavior},
volume = {1},
number = {1},
pages = {3-14},
doi = {https://doi.org/10.1002/j.2162-6057.1967.tb00002.x},
url = {https://onlinelibrary.wiley.com/doi/abs/10.1002/j.2162-6057.1967.tb00002.x},
year = {1967}
}

@article{GUZIK2023100065,
title = {The originality of machines: AI takes the Torrance Test},
journal = {Journal of Creativity},
volume = {33},
number = {3},
pages = {100065},
year = {2023},
issn = {2713-3745},
doi = {https://doi.org/10.1016/j.yjoc.2023.100065},
url = {https://www.sciencedirect.com/science/article/pii/S2713374523000249},
author = {Erik E. Guzik and Christian Byrge and Christian Gilde}
}

@article{HAASE_2023,
title = {Artificial muses: Generative artificial intelligence chatbots have risen to human-level creativity},
journal = {Journal of Creativity},
volume = {33},
number = {3},
pages = {100066},
year = {2023},
issn = {2713-3745},
doi = {https://doi.org/10.1016/j.yjoc.2023.100066},
url = {https://www.sciencedirect.com/science/article/pii/S2713374523000250},
author = {Jennifer Haase and Paul H.P. Hanel}
}

@inproceedings{Chakrabarty_CSTsLLM,
author = {Chakrabarty, Tuhin and Padmakumar, Vishakh and Brahman, Faeze and Muresan, Smaranda},
title = {Creativity Support in the Age of Large Language Models: An Empirical Study Involving Professional Writers},
year = {2024},
isbn = {9798400704857},
publisher = {Association for Computing Machinery},
address = {New York, NY, USA},
url = {https://doi.org/10.1145/3635636.3656201},
doi = {10.1145/3635636.3656201},
booktitle = {Proceedings of the 16th Conference on Creativity \& Cognition},
pages = {132–155},
numpages = {24},
location = {Chicago, IL, USA},
series = {C\&C '24}
}

@article{hills_entropy_2025,
	title = {An entropy modulation theory of creative exploration.},
	volume = {132},
	issn = {1939-1471, 0033-295X},
	url = {https://doi.apa.org/doi/10.1037/rev0000511},
	doi = {10.1037/rev0000511},
	language = {en},
	number = {1},
	urldate = {2025-07-04},
	journal = {Psychological Review},
	author = {Hills, Thomas T. and Kenett, Yoed N.},
	month = jan,
	year = {2025},
	pages = {239--251},
}

@article{hubert_current_2024,
	title = {The current state of artificial intelligence generative language models is more creative than humans on divergent thinking tasks},
	volume = {14},
	issn = {2045-2322},
	url = {https://doi.org/10.1038/s41598-024-53303-w},
	doi = {10.1038/s41598-024-53303-w},
	number = {1},
	journal = {Scientific Reports},
	author = {Hubert, Kent F. and Awa, Kim N. and Zabelina, Darya L.},
	month = feb,
	year = {2024},
	pages = {3440},
}

@article{grassini_artificial_2025,
	title = {Artificial {Creativity}? {Evaluating} {AI} {Against} {Human} {Performance} in {Creative} {Interpretation} of {Visual} {Stimuli}},
	volume = {41},
	issn = {1044-7318},
	url = {https://doi.org/10.1080/10447318.2024.2345430},
	doi = {10.1080/10447318.2024.2345430},
	number = {7},
	journal = {International Journal of Human–Computer Interaction},
	author = {Grassini, Simone and Koivisto, Mika},
	month = apr,
	year = {2025},
	note = {Publisher: Taylor \& Francis},
	pages = {4037--4048},
}

@article{kenett_semantic_2019,
	title = {A {Semantic} {Network} {Cartography} of the {Creative} {Mind}},
	volume = {23},
	issn = {1364-6613},
	url = {https://doi.org/10.1016/j.tics.2019.01.007},
	doi = {10.1016/j.tics.2019.01.007},
	number = {4},
	urldate = {2025-08-16},
	journal = {Trends in Cognitive Sciences},
	author = {Kenett, Yoed N. and Faust, Miriam},
	month = apr,
	year = {2019},
	note = {Publisher: Elsevier},
	pages = {271--274},
}

@article{silvia_assessing_2008,
	title = {Assessing {Creativity} {With} {Divergent} {Thinking} {Tasks}: {Exploring} the {Reliability} and {Validity} of {New} {Subjective} {Scoring} {Methods}},
	volume = {2},
	shorttitle = {Assessing {Creativity} {With} {Divergent} {Thinking} {Tasks}},
	doi = {10.1037/1931-3896.2.2.68},
	journal = {Psychology of Aesthetics, Creativity, and the Arts},
	author = {Silvia, Paul and Winterstein, Beate and Willse, John and Barona, Christopher and Cram, Joshua and Hess, Karl and Martinez, Jenna and Richard, Crystal},
	month = may,
	year = {2008},
	pages = {68--85},
}

@inproceedings{Anderson,
author = {Anderson, Barrett R and Shah, Jash Hemant and Kreminski, Max},
title = {Homogenization Effects of Large Language Models on Human Creative Ideation},
year = {2024},
isbn = {9798400704857},
publisher = {Association for Computing Machinery},
address = {New York, NY, USA},
url = {https://doi.org/10.1145/3635636.3656204},
doi = {10.1145/3635636.3656204},
booktitle = {Proceedings of the 16th Conference on Creativity \& Cognition},
pages = {413–425},
numpages = {13},
location = {Chicago, IL, USA},
series = {C\&C '24}
}

@article{Kenett_percolation,
author = {Yoed N. Kenett  and Orr Levy  and Dror Y. Kenett  and H. Eugene Stanley  and Miriam Faust  and Shlomo Havlin },
title = {Flexibility of thought in high creative individuals represented by percolation analysis},
journal = {Proceedings of the National Academy of Sciences},
volume = {115},
number = {5},
pages = {867-872},
year = {2018},
doi = {10.1073/pnas.1717362115},
URL = {https://www.pnas.org/doi/abs/10.1073/pnas.1717362115}}

@article{koivisto_best_2023,
	title = {Best humans still outperform artificial intelligence in a creative divergent thinking task},
	volume = {13},
	issn = {2045-2322},
	url = {https://doi.org/10.1038/s41598-023-40858-3},
	doi = {10.1038/s41598-023-40858-3},
	number = {1},
	journal = {Scientific Reports},
	author = {Koivisto, Mika and Grassini, Simone},
	month = sep,
	year = {2023},
	pages = {13601},
}

@article{Li03072021,
author = {Yangping Li and Yoed N. Kenett and Weiping Hu and Roger E. Beaty},
title = {Flexible Semantic Network Structure Supports the Production of Creative Metaphor},
journal = {Creativity Research Journal},
volume = {33},
number = {3},
pages = {209--223},
year = {2021},
publisher = {Routledge},
doi = {10.1080/10400419.2021.1879508},


URL = { 
    
        https://doi.org/10.1080/10400419.2021.1879508

    
    

}

}

@article{tumminello_tool_2005,
	title = {A tool for filtering information in complex systems},
	volume = {102},
	url = {https://doi.org/10.1073/pnas.0500298102},
	doi = {10.1073/pnas.0500298102},
	number = {30},
	urldate = {2025-11-15},
	journal = {Proceedings of the National Academy of Sciences},
	author = {Tumminello, M. and Aste, T. and Di Matteo, T. and Mantegna, R. N.},
	month = jul,
	year = {2005},
	note = {Publisher: Proceedings of the National Academy of Sciences},
	pages = {10421--10426},
}

@article{kenett_PMFG_2013,
	title = {Semantic organization in children with cochlear implants: computational analysis of verbal fluency},
	volume = {Volume 4 - 2013},
	issn = {1664-1078},
	url = {https://www.frontiersin.org/journals/psychology/articles/10.3389/fpsyg.2013.00543},
	doi = {10.3389/fpsyg.2013.00543},
	journal = {Frontiers in Psychology},
	author = {Kenett, Yoed N. and Wechsler-Kashi, Deena and Kenett, Dror Y. and Schwartz, Richard G. and Ben Jacob, Eshel and Faust, Miriam},
	year = {2013},
}

@article{christensen_remotely_2018,
	title = {Remotely {Close} {Associations}: {Openness} to {Experience} and {Semantic} {Memory} {Structure}},
	volume = {32},
	issn = {1099-0984},
	shorttitle = {Remotely {Close} {Associations}},
	url = {https://onlinelibrary.wiley.com/doi/abs/10.1002/per.2157},
	doi = {10.1002/per.2157},
	language = {en},
	number = {4},
	urldate = {2025-10-08},
	journal = {European Journal of Personality},
	author = {Christensen, Alexander P. and Kenett, Yoed N. and Cotter, Katherine N. and Beaty, Roger E. and Silvia, Paul J.},
	year = {2018},
	note = {\_eprint: https://onlinelibrary.wiley.com/doi/pdf/10.1002/per.2157},
	pages = {480--492},

}

@inproceedings{Cremaschi_2025,
author = {Cremaschi, Michele and Grigis, Paolo and Dorfmann, Max and De Angeli, Antonella},
title = {Exposing the Ideology of Large Language Models with Creative Practices},
year = {2025},
isbn = {9798400714856},
publisher = {Association for Computing Machinery},
address = {New York, NY, USA},
url = {https://doi.org/10.1145/3715336.3735715},
doi = {10.1145/3715336.3735715},
booktitle = {Proceedings of the 2025 ACM Designing Interactive Systems Conference},
pages = {1928–1940},
numpages = {13},
location = {
},
series = {DIS '25}
}

@article{aru_artificial_2025,
	title = {Artificial {Intelligence} and the {Internal} {Processes} of {Creativity}},
	volume = {59},
	issn = {0022-0175},
	url = {https://doi.org/10.1002/jocb.1530},
	doi = {10.1002/jocb.1530},
	number = {2},
	urldate = {2025-09-10},
	journal = {The Journal of Creative Behavior},
	author = {Aru, Jaan},
	month = jun,
	year = {2025},
	note = {Publisher: John Wiley \& Sons, Ltd},
	pages = {e1530},
}

@article{massara_network_2016,
	title = {Network {Filtering} for {Big} {Data}: {Triangulated} {Maximally} {Filtered} {Graph}},
	volume = {5},
	issn = {2051-1310},
	url = {https://doi.org/10.1093/comnet/cnw015},
	doi = {10.1093/comnet/cnw015},
	number = {2},
	journal = {Journal of Complex Networks},
	author = {Massara, Guido Previde and Di Matteo, T. and Aste, Tomaso},
	month = jun,
	year = {2016},
	pages = {161--178},
}

@article{nelson_what_2000,
	title = {What is free association and what does it measure?},
	volume = {28},
	issn = {1532-5946},
	url = {https://doi.org/10.3758/BF03209337},
	doi = {10.3758/BF03209337},
	number = {6},
	journal = {Memory \& Cognition},
	author = {Nelson, Douglas L. and Mcevoy, Cathy L. and Dennis, Simon},
	month = nov,
	year = {2000},
	pages = {887--899},
}

@article{olson_naming_2021,
	title = {Naming unrelated words predicts creativity},
	volume = {118},
	issn = {0027-8424, 1091-6490},
	url = {https://pnas.org/doi/full/10.1073/pnas.2022340118},
	doi = {10.1073/pnas.2022340118},
	language = {en},
	number = {25},
	urldate = {2025-01-13},
	journal = {Proceedings of the National Academy of Sciences},
	author = {Olson, Jay A. and Nahas, Johnny and Chmoulevitch, Denis and Cropper, Simon J. and Webb, Margaret E.},
	month = jun,
	year = {2021},
	pages = {e2022340118},
}

@article{Oppenlaender18082025,
author = {Jonas Oppenlaender and Rhema Linder and Johanna Silvennoinen},
title = {Prompting AI Art: An Investigation into the Creative Skill of Prompt Engineering},
journal = {International Journal of Human–Computer Interaction},
volume = {41},
number = {16},
pages = {10207--10229},
year = {2025},
publisher = {Taylor \& Francis},
doi = {10.1080/10447318.2024.2431761},


URL = { 
    
        https://doi.org/10.1080/10447318.2024.2431761
}

}

@inproceedings{Chakrabarty,
author = {Chakrabarty, Tuhin and Laban, Philippe and Agarwal, Divyansh and Muresan, Smaranda and Wu, Chien-Sheng},
title = {Art or Artifice? Large Language Models and the False Promise of Creativity},
year = {2024},
isbn = {9798400703300},
publisher = {Association for Computing Machinery},
address = {New York, NY, USA},
url = {https://doi.org/10.1145/3613904.3642731},
doi = {10.1145/3613904.3642731},
booktitle = {Proceedings of the 2024 CHI Conference on Human Factors in Computing Systems},
articleno = {30},
numpages = {34},
location = {Honolulu, HI, USA},
series = {CHI '24},
}

@inproceedings{Rajcic,
author = {Rajcic, Nina and Llano Rodriguez, Maria Teresa and McCormack, Jon},
title = {Towards a Diffractive Analysis of Prompt-Based Generative AI},
year = {2024},
isbn = {9798400703300},
publisher = {Association for Computing Machinery},
address = {New York, NY, USA},
url = {https://doi.org/10.1145/3613904.3641971},
doi = {10.1145/3613904.3641971},
booktitle = {Proceedings of the 2024 CHI Conference on Human Factors in Computing Systems},
articleno = {844},
numpages = {15},
location = {Honolulu, HI, USA},
series = {CHI '24}
}

@inproceedings{Tankelevitch,
author = {Tankelevitch, Lev and Kewenig, Viktor and Simkute, Auste and Scott, Ava Elizabeth and Sarkar, Advait and Sellen, Abigail and Rintel, Sean},
title = {The Metacognitive Demands and Opportunities of Generative AI},
year = {2024},
isbn = {9798400703300},
publisher = {Association for Computing Machinery},
address = {New York, NY, USA},
url = {https://doi.org/10.1145/3613904.3642902},
doi = {10.1145/3613904.3642902},
booktitle = {Proceedings of the 2024 CHI Conference on Human Factors in Computing Systems},
articleno = {680},
numpages = {24},
location = {Honolulu, HI, USA},
series = {CHI '24}
}

@inproceedings{Rayan,
author = {Rayan, Jude and Kanetkar, Dhruv and Gong, Yifan and Yang, Yuewen and Palani, Srishti and Xia, Haijun and Dow, Steven P.},
title = {Exploring the Potential for Generative AI-based Conversational Cues for Real-Time Collaborative Ideation},
year = {2024},
isbn = {9798400704857},
publisher = {Association for Computing Machinery},
address = {New York, NY, USA},
url = {https://doi.org/10.1145/3635636.3656184},
doi = {10.1145/3635636.3656184},
booktitle = {Proceedings of the 16th Conference on Creativity \& Cognition},
pages = {117–131},
numpages = {15},
location = {Chicago, IL, USA},
series = {C\&C '24}
}

@inproceedings{Chang,
author = {Chang, Minsuk and Druga, Stefania and Fiannaca, Alexander J. and Vergani, Pedro and Kulkarni, Chinmay and Cai, Carrie J and Terry, Michael},
title = {The Prompt Artists},
year = {2023},
isbn = {9798400701801},
publisher = {Association for Computing Machinery},
address = {New York, NY, USA},
url = {https://doi.org/10.1145/3591196.3593515},
doi = {10.1145/3591196.3593515},
booktitle = {Proceedings of the 15th Conference on Creativity and Cognition},
pages = {75–87},
numpages = {13},
location = {Virtual Event, USA},
series = {C\&C '23}
}

@article{organisciak_ocsai,
title = {Beyond semantic distance: Automated scoring of divergent thinking greatly improves with large language models},
journal = {Thinking Skills and Creativity},
volume = {49},
pages = {101356},
year = {2023},
issn = {1871-1871},
doi = {https://doi.org/10.1016/j.tsc.2023.101356},
url = {https://www.sciencedirect.com/science/article/pii/S1871187123001256},
author = {Peter Organisciak and Selcuk Acar and Denis Dumas and Kelly Berthiaume}
}

@incollection{geroimenko_hall_2025,
	address = {Cham},
	title = {Generative {AI} {Hallucinations} in {Healthcare}: {A} {Challenge} for {Prompt} {Engineering} and {Creativity}},
	isbn = {978-3-031-86551-0},
	url = {https://doi.org/10.1007/978-3-031-86551-0_17},
	doi = {10.1007/978-3-031-86551-0_17},
	booktitle = {Human-{Computer} {Creativity}: {Generative} {AI} in {Education}, {Art}, and {Healthcare}},
	publisher = {Springer Nature Switzerland},
	author = {Geroimenko, Vladimir},
	editor = {Geroimenko, Vladimir},
	year = {2025},
	pages = {321--335},
}

@article{campidelli_creativity_2026,
	title = {Creativity, the fountain of youth: {Association} between creativity and semantic memory networks across the lifespan},
	volume = {266},
	issn = {0010-0277},
	url = {https://www.sciencedirect.com/science/article/pii/S0010027725002598},
	doi = {https://doi.org/10.1016/j.cognition.2025.106318},
	journal = {Cognition},
	author = {Campidelli, Lorenzo and Domanti, Umberto and Fusi, Giulia and Kenett, Yoed N. and Agnoli, Sergio},
	year = {2026},
	pages = {106318},
}

@book{torrance_1966,
  title={Torrance Tests of Creative Thinking: Norms-technical Manual. Research Edition. Verbal Tests, Forms A and B. Figural Tests, Forms A and B.},
  author={Torrance, Ellis Paul},
  year={1966},
  publisher={Personnel Press}
}

@article{rafner_creativity_2023,
	title = {Creativity in the age of generative {AI}},
	volume = {7},
	issn = {2397-3374},
	url = {https://doi.org/10.1038/s41562-023-01751-1},
	doi = {10.1038/s41562-023-01751-1},
	number = {11},
	journal = {Nature Human Behaviour},
	author = {Rafner, Janet and Beaty, Roger E. and Kaufman, James C. and Lubart, Todd and Sherson, Jacob},
	month = nov,
	year = {2023},
	pages = {1836--1838},
}

@inproceedings{stevenson_2022putting,
  author = {Claire Stevenson and Iris Smal and Matthijs Baas and Raoul Grasman and Han van der Maas},
  title = {Putting GPT-3's Creativity to the (Alternative Uses) Test},
  pages = {164-168},
  booktitle = {Proceedings of the 13th International Conference on Computational Creativity},
  editor = {Maria M. Hedblom and Anna Aurora Kantosalo and Roberto Confalonieri and Oliver Kutz and Tony Veale},
  year = {2022},
  address = {Bozen-Bolzano, Italy},
  isbn = {978-989-54160-4-2},
  publisher = {Association for Computational Creativity},
  url = {http://computationalcreativity.net/iccc22/papers/ICCC-2022_paper_140.pdf}
}

@inproceedings{goes_2023pushing,
  author = {Fabr{\'{i}}cio G{\'{o}}es and Piotr Sawicki and Marek Grzes and Marco Volpe and Jacob Watson},
  title = {Pushing GPT's Creativity to Its Limits: Alternative Uses and Torrance Tests},
  pages = {342-346},
  booktitle = {Proceedings of the 14th International Conference on Computational Creativity},
  editor = {Alison Pease and Jo{\~{a}}o Miguel Cunha and Maya Ackerman and Daniel G. Brown},
  year = {2023},
  address = {Ontario, Canada},
  isbn = {978-989-54160-5-9},
  publisher = {Association for Computational Creativity},
  url = {https://computationalcreativity.net/iccc23/papers/ICCC-2023_paper_90.pdf}
}

@inproceedings{peeperkorn_temperature_2024,
  author = {Max Peeperkorn and Tom Kouwenhoven and Dan Brown and Anna Jordanous},
  title = {Is Temperature the Creativity Parameter of Large Language Models?},
  pages = {226-235},
  booktitle = {Proceedings of the 15th International Conference on Computational Creativity},
  editor = {Kazjon Grace and Maria Teresa Llano and Pedro Martins and Maria M. Hedblom},
  year = {2024},
  address = {J{\"o}nk{\"o}ping, Sweden},
  isbn = {978-989-54160-6-6},
  publisher = {Association for Computational Creativity},
  url = {https://computationalcreativity.net/iccc24/papers/ICCC24_paper_70.pdf}
}

@article{Rhodes,
 ISSN = {00317217},
 URL = {http://www.jstor.org/stable/20342603},
 author = {Mel Rhodes},
 journal = {The Phi Delta Kappan},
 number = {7},
 pages = {305--310},
 publisher = {Phi Delta Kappa International},
 title = {An Analysis of Creativity},
 urldate = {2025-08-16},
 volume = {42},
 year = {1961}
}

@article{Shneiderman,
author = {Shneiderman, Ben},
title = {Creativity support tools},
year = {2002},
issue_date = {October 2002},
publisher = {Association for Computing Machinery},
address = {New York, NY, USA},
volume = {45},
number = {10},
issn = {0001-0782},
url = {https://doi.org/10.1145/570907.570945},
doi = {10.1145/570907.570945},
journal = {Commun. ACM},
month = oct,
pages = {116–120},
numpages = {5}
}

@article{RUNCO_2023,
title = {AI can only produce artificial creativity},
journal = {Journal of Creativity},
volume = {33},
number = {3},
pages = {100063},
year = {2023},
issn = {2713-3745},
doi = {https://doi.org/10.1016/j.yjoc.2023.100063},
url = {https://www.sciencedirect.com/science/article/pii/S2713374523000225},
author = {Mark A. Runco}
}

@article{Runco_2025,
author = {Mark A. Runco},
title = {Updating the Standard Definition of Creativity to Account for the Artificial Creativity of AI},
journal = {Creativity Research Journal},
volume = {37},
number = {1},
pages = {1--5},
year = {2025},
publisher = {Routledge},
doi = {10.1080/10400419.2023.2257977},    

}

@article{Runco_2012,
author = {Mark A. Runco and Garrett J. Jaeger},
title = {The Standard Definition of Creativity},
journal = {Creativity Research Journal},
volume = {24},
number = {1},
pages = {92--96},
year = {2012},
publisher = {Routledge},
doi = {10.1080/10400419.2012.650092},
}

@article{said-metwaly_divergent_2024,
	title = {Divergent thinking and creative achievement—{How} strong is the link? {An} updated meta-analysis.},
	volume = {18},
	issn = {1931-390X(Electronic),1931-3896(Print)},
	doi = {10.1037/aca0000507},
	number = {5},
	journal = {Psychology of Aesthetics, Creativity, and the Arts},
	author = {Said-Metwaly, Sameh and Taylor, Christa L. and Camarda, Anaëlle and Barbot, Baptiste},
	year = {2024},
	place = {US},
Publisher = {Educational Publishing Foundation},
	pages = {869--881},
}

@article{shanahan_role_2023,
	title = {Role play with large language models},
	volume = {623},
	issn = {1476-4687},
	url = {https://doi.org/10.1038/s41586-023-06647-8},
	doi = {10.1038/s41586-023-06647-8},
	number = {7987},
	journal = {Nature},
	author = {Shanahan, Murray and McDonell, Kyle and Reynolds, Laria},
	month = nov,
	year = {2023},
	pages = {493--498},
}

@article{ovando-tellez_brain_nodate,
	title = {Brain connectivity–based prediction of real-life creativity is mediated by semantic memory structure},
	volume = {8},
        year={2022},
	url = {https://doi.org/10.1126/sciadv.abl4294},
	doi = {10.1126/sciadv.abl4294},
	number = {5},
	urldate = {2025-08-18},
	journal = {Science Advances},
	author = {Ovando-Tellez, Marcela and Kenett, Yoed N. and Benedek, Mathias and Bernard, Matthieu and Belo, Joan and Beranger, Benoit and Bieth, Theophile and Volle, Emmanuelle},
	note = {Publisher: American Association for the Advancement of Science},
	pages = {eabl4294},
}

@article{he_relation_2020,
	title = {The relation between semantic memory structure, associative abilities, and verbal and figural creativity},
	volume = {27},
	issn = {1354-6783},
	url = {https://doi.org/10.1080/13546783.2020.1819415},
	doi = {10.1080/13546783.2020.1819415},
	number = {2},
	journal = {Thinking \& Reasoning},
	author = {He, Li and Kenett, Yoed N. and Zhuang, Kaixiang and Liu, Cheng and Zeng, Rongcan and Yan, Tingrui and Huo, Tengbin and Qiu, Jiang},
	month = sep,
	year = {2020},
	note = {Publisher: Routledge},
	pages = {268--293},
}

@article{benedek_how_2017,
	title = {How semantic memory structure and intelligence contribute to creative thought: a network science approach},
	volume = {23},
	issn = {1354-6783},
	url = {https://doi.org/10.1080/13546783.2016.1278034},
	doi = {10.1080/13546783.2016.1278034},
	number = {2},
	journal = {Thinking \& Reasoning},
	author = {Benedek, Mathias and Kenett, Yoed N. and Umdasch, Konstantin and Anaki, David and Faust, Miriam and Neubauer, Aljoscha C.},
	month = apr,
	year = {2017},
	note = {Publisher: Routledge},
	pages = {158--183},
}

@article{Siew,
author = {Siew, Cynthia S. Q. and Wulff, Dirk U. and Beckage, Nicole M. and Kenett, Yoed N.},
title = {Cognitive Network Science: A Review of Research on Cognition through the Lens of Network Representations, Processes, and Dynamics},
journal = {Complexity},
volume = {2019},
number = {1},
pages = {2108423},
doi = {https://doi.org/10.1155/2019/2108423},
url = {https://onlinelibrary.wiley.com/doi/abs/10.1155/2019/2108423},
year = {2019}
}

@article{taylor_artificial_2021,
	title = {Artificial cognition: {How} experimental psychology can help generate explainable artificial intelligence},
	volume = {28},
	issn = {1531-5320},
	url = {https://doi.org/10.3758/s13423-020-01825-5},
	doi = {10.3758/s13423-020-01825-5},
	number = {2},
	journal = {Psychonomic Bulletin \& Review},
	author = {Taylor, J. Eric T. and Taylor, Graham W.},
	month = apr,
	year = {2021},
	pages = {454--475},
}

@Article{Vinchon,
AUTHOR = {Vinchon, Florent and Gironnay, Valentin and Lubart, Todd},
TITLE = {GenAI Creativity in Narrative Tasks: Exploring New Forms of Creativity},
JOURNAL = {Journal of Intelligence},
VOLUME = {12},
YEAR = {2024},
NUMBER = {12},
ARTICLE-NUMBER = {125},
URL = {https://www.mdpi.com/2079-3200/12/12/125},
PubMedID = {39728093},
ISSN = {2079-3200},
DOI = {10.3390/jintelligence12120125}
}

@article{WANG2025100103,
title = {The fluency-based semantic network of LLMs differs from humans},
journal = {Computers in Human Behavior: Artificial Humans},
volume = {3},
pages = {100103},
year = {2025},
issn = {2949-8821},
doi = {https://doi.org/10.1016/j.chbah.2024.100103},
url = {https://www.sciencedirect.com/science/article/pii/S294988212400063X},
author = {Ye Wang and Yaling Deng and Ge Wang and Tong Li and Hongjiang Xiao and Yuan Zhang}
}

@article{zemla_snafu_2020,
	title = {{SNAFU}: {The} {Semantic} {Network} and {Fluency} {Utility}},
	volume = {52},
	issn = {1554-3528},
	url = {https://doi.org/10.3758/s13428-019-01343-w},
	doi = {10.3758/s13428-019-01343-w},
	number = {4},
	journal = {Behavior Research Methods},
	author = {Zemla, Jeffrey C. and Cao, Kesong and Mueller, Kimberly D. and Austerweil, Joseph L.},
	month = aug,
	year = {2020},
	pages = {1681--1699},
}

@inproceedings{Ghajargar,
author = {Ghajargar, Maliheh and Bardzell, Jeffrey and Lagerkvist, Love},
title = {A Redhead Walks into a Bar: Experiences of Writing Fiction with Artificial Intelligence},
year = {2022},
isbn = {9781450399555},
publisher = {Association for Computing Machinery},
address = {New York, NY, USA},
url = {https://doi.org/10.1145/3569219.3569418},
doi = {10.1145/3569219.3569418},
booktitle = {Proceedings of the 25th International Academic Mindtrek Conference},
pages = {230–241},
numpages = {12},
location = {Tampere, Finland},
series = {Academic Mindtrek '22}
}

@article{kenett_network,
	title = {Investigating the structure of semantic networks in low and high creative persons},
	volume = {Volume 8 - 2014},
	issn = {1662-5161},
	url = {https://www.frontiersin.org/journals/human-neuroscience/articles/10.3389/fnhum.2014.00407},
	doi = {10.3389/fnhum.2014.00407},
	journal = {Frontiers in Human Neuroscience},
	author = {Kenett, Yoed N. and Anaki, David and Faust, Miriam},
	year = {2014},
}

@article{Jauk,
author = {Emanuel Jauk and Mathias Benedek and Aljoscha C. Neubauer},
title ={The Road to Creative Achievement: A Latent Variable Model of Ability and Personality Predictors},

journal = {European Journal of Personality},
volume = {28},
number = {1},
pages = {95-105},
year = {2014},
doi = {10.1002/per.1941},
    note ={PMID: 24532953},
URL = {https://doi.org/10.1002/per.1941},
}

@article{vinchon_manifesto_2023,
	title = {Artificial {Intelligence} \& {Creativity}: {A} {Manifesto} for {Collaboration}},
	volume = {57},
	issn = {0022-0175},
	url = {https://doi.org/10.1002/jocb.597},
	doi = {10.1002/jocb.597},
	number = {4},
	urldate = {2025-08-28},
	journal = {The Journal of Creative Behavior},
	author = {Vinchon, Florent and Lubart, Todd and Bartolotta, Sabrina and Gironnay, Valentin and Botella, Marion and Bourgeois-Bougrine, Samira and Burkhardt, Jean-Marie and Bonnardel, Nathalie and Corazza, Giovanni Emanuele and Glăveanu, Vlad and Hanchett Hanson, Michael and Ivcevic, Zorana and Karwowski, Maciej and Kaufman, James C. and Okada, Takeshi and Reiter-Palmon, Roni and Gaggioli, Andrea},
	month = dec,
	year = {2023},
	note = {Publisher: John Wiley \& Sons, Ltd},
	pages = {472--484},
}

@article{amabile_dynamic_2016,
	title = {The dynamic componential model of creativity and innovation in organizations: {Making} progress, making meaning},
	volume = {36},
	issn = {0191-3085},
	url = {https://www.sciencedirect.com/science/article/pii/S0191308516300053},
	doi = {https://doi.org/10.1016/j.riob.2016.10.001},
	journal = {Research in Organizational Behavior},
	author = {Amabile, Teresa M. and Pratt, Michael G.},
	year = {2016},
	pages = {157--183},
}

@incollection{karwowski_chapter_2017,
	address = {San Diego},
	series = {Explorations in {Creativity} {Research}},
	title = {Chapter 18 - {The} {Dynamic} {Force} {Before} {Intrinsic} {Motivation}: {Exploring} {Creative} {Needs}},
	isbn = {978-0-12-809790-8},
	url = {https://www.sciencedirect.com/science/article/pii/B9780128097908000182},
	booktitle = {The {Creative} {Self}},
	publisher = {Academic Press},
	author = {Luria, Sarah R. and Kaufman, James C.},
	editor = {Karwowski, Maciej and Kaufman, James C.},
	year = {2017},
	doi = {https://doi.org/10.1016/B978-0-12-809790-8.00018-2},
	pages = {317--325},
}

@article{cropley_averagecreativity_2025,
	title = {“{The} {Cat} {Sat} on the …?” {Why} {Generative} {AI} {Has} {Limited} {Creativity}},
	volume = {59},
	issn = {0022-0175, 2162-6057},
	shorttitle = {“{The} {Cat} {Sat} on the …?},
	url = {https://onlinelibrary.wiley.com/doi/10.1002/jocb.70077},
	doi = {10.1002/jocb.70077},
	language = {en},
	number = {4},
	urldate = {2025-11-16},
	journal = {The Journal of Creative Behavior},
	author = {Cropley, David H.},
	month = dec,
	year = {2025},
	pages = {e70077},
}

@inproceedings{deangeli_2025,
author = {De Angeli, Antonella and Grigis, Paolo},
title = {Setting the Stage for Fictional and Artificial Characters},
year = {2025},
isbn = {9798400721021},
publisher = {Association for Computing Machinery},
address = {New York, NY, USA},
url = {https://doi.org/10.1145/3750069.3750155},
doi = {10.1145/3750069.3750155},
booktitle = {Proceedings of the 16th Biannual Conference of the Italian SIGCHI Chapter},
articleno = {19},
numpages = {10},
location = {
},
series = {CHItaly '25}
}

@article{beghetto_creative_2014,
	title = {Creative mortification: {An} initial exploration.},
	volume = {8},
	issn = {1931-390X(Electronic),1931-3896(Print)},
	doi = {10.1037/a0036618},
	number = {3},
	journal = {Psychology of Aesthetics, Creativity, and the Arts},
	author = {Beghetto, Ronald A.},
	year = {2014},
Place = {US},
Publisher = {Educational Publishing Foundation},
	pages = {266--276},
}

@inproceedings{Cremaschi_steamp,
author = {Cremaschi, Michele and Dorfmann, Max and De Angeli, Antonella},
title = {A Steampunk Critique of Machine Learning Acceleration},
year = {2024},
isbn = {9798400705830},
publisher = {Association for Computing Machinery},
address = {New York, NY, USA},
url = {https://doi.org/10.1145/3643834.3660688},
doi = {10.1145/3643834.3660688},
booktitle = {Proceedings of the 2024 ACM Designing Interactive Systems Conference},
pages = {246–257},
numpages = {12},
location = {Copenhagen, Denmark},
series = {DIS '24}
}

@article{cosgrove_age_2023,
	title = {Age differences in semantic network structure: {Acquiring} knowledge shapes semantic memory},
	volume = {38},
	issn = {1939-1498},
	shorttitle = {Age differences in semantic network structure},
	doi = {10.1037/pag0000721},
	number = {2},
	journal = {Psychology and Aging},
	author = {Cosgrove, Abigail L. and Beaty, Roger E. and Diaz, Michele T. and Kenett, Yoed N.},
	year = {2023},
Place = {US},
Publisher = {American Psychological Association},
	pages = {87--102},
}

@article{cosgrove_associative_nodate,
	title = {Associative {Thinking} and {Creative} {Ability} in {Older} {Adulthood}},
	issn = {1040-0419},
	url = {https://doi.org/10.1080/10400419.2024.2443720},
	doi = {10.1080/10400419.2024.2443720},
	journal = {Creativity Research Journal},
	author = {Cosgrove, Abigail L. and Diaz, Michele T. and DiStefano, Paul V. and Kenett, Yoed N. and Beaty, Roger E.},
	note = {Publisher: Routledge},
year = {2025},
	pages = {1--14},
}


\appendix
\appendix
\section{Human instructions and ChatGPT-4o prompts}\label{sec: appendixA} 
Table \ref{table_appendix} presents a comparison between the instructions given to human participants and the prompts received by ChatGPT-4o across all tasks\footnote{The instructions are translated from Italian to English}. Human participants saw the target word (e.g., animals) printed on a sheet of paper, and in the examples reported in Table \ref{table_appendix} and in the Appendix \ref{appendix_human}, this appears in square brackets. For obtaining machine data, no system prompts were used and one example was included in the instructions (i.e., a one-shot prompting strategy was applied \cite{brown_language_2020}, see Table \ref{table_appendix}). For each ChatGPT-4o participant, the chat window was not reset between tasks. 
\begin{table*}
  \centering
  \caption{Comparison between human instructions and ChatGPT-4o prompts}
  \Description{Table 7 presents a comparison between the human instructions and the prompts given to ChatGPT-4o across all experimental tasks (Verbal Fluency Test, Free Association Task, Alternate Uses Task).}
  \label{table_appendix}
  \begin{tabular}{p{7.7cm} p{7.7cm}}
    \toprule
    \textbf{Human instructions} & \textbf{ChatGPT-4o prompts} \\
    \midrule

    \textbf{Verbal Fluency Test 1} \newline
    Now I will present you with a sheet of paper on which a word representing a category [animals] is written, and I will ask you to tell me as many words that come to your mind that belong to that particular category. You will have one minute. For example, for the category Trees, you might say: Oak, Poplar, Fir, etc.
     &
    \textbf{Verbal Fluency Test 1} \newline
    Now I will present you with a word representing a category, and I will ask you to tell me as many words that come to your mind that belong to that particular category. You will have one minute. For example, for the category Trees, you might say: Oak, Poplar, Fir, etc. The word is "animals". \\

    \textbf{Verbal Fluency Test 2} \newline
    Now, the category is [fruit and vegetables]. You will have one minute.
    &
    \textbf{Verbal Fluency Test 2} \newline
    Now, the category is "fruit and vegetables". You will have one minute. \\
    \midrule

    \textbf{Free Association Task 1} \newline
    Now I will present you with a sheet of paper on which a word [building] will be written, and I will ask you to tell me as many words as come to your mind that are associated with the presented word. You will have one minute. For example, for the word Father, you might say: Mother, Son, Family, etc.
    &
    \textbf{Free Association Task 1} \newline
    Now I will present you a word, and I will ask you to tell me as many words as come to your mind that are associated with the presented word. You will have one minute. For example, for the word Father, you might say: Mother, Son, Family, etc. The word is "building". \\

    \textbf{Free Association Task 2} \newline
    Now, the word is [notebook]. You will have one minute.
    &
    \textbf{Free Association Task 2} \newline
    Now, the word is "notebook". You will have one minute. \\
    \midrule

    \textbf{Alternate Uses Task 1} \newline
    Now I will present you with a sheet of paper on which a word [shoe] denoting an object is written. The task consists of producing original uses for that object. All objects have a common use. For example, what is the common use of a tablecloth? To put it on the table when we set it, in order to avoid getting it dirty. But that is the common use; what we are interested in are the original uses of the object, which I will tell you in a moment. For instance, if we take the tablecloth off the table and put it on ourselves, it can become a blanket if we are cold or a cape if we are playing with a child. Or, if it is old, we could cut it up and use it as rags. Your task is to come up with as many original alternative uses for that object as possible. You will have three minutes.
    &
    \textbf{Alternate Uses Task 1} \newline
   Now I will present you with a word denoting an object. The task consists of producing original uses for that object. All objects have a common use. For example, what is the common use of a tablecloth? To put it on the table when we set it, in order to avoid getting it dirty. But that is the common use; what we are interested in are the original uses of the object, which I will tell you in a moment. For instance, if we take the tablecloth off the table and put it on ourselves, it can become a blanket if we are cold or a cape if we are playing with a child. Or, if it is old, we could cut it up and use it as rags. Your task is to come up with as many original alternative uses for that object as possible. You will have three minutes. The object is ``shoe''. \\
 
    \textbf{Alternate Uses Task 2} \newline
    Now, the object is [fork]. You will have three minutes.
    &
    \textbf{Alternate Uses Task 2} \newline
    Now, the object is "fork". You will have three minutes. \\
    \bottomrule
  \end{tabular}
\end{table*}

\section{Example of Human and ChatGPT-4o performances}\label{sec: appendixB}
This Appendix is organised as follows. Section \ref{Human} presents the interaction (i.e., instructions and outputs) for a human participant. Section \ref{AI} presents the interaction (i.e., prompts and outputs) for a ChatGPT-4o participant. To anonymise the responses for the expert coders, in the final database, all hyphens in ChatGPT-4o AUT responses were replaced with commas. Moreover, any sentences appearing before and after the list of words related to the tasks were excluded (e.g., here is a list of animals that come to mind or time is up!). All responses provided by human and ChatGPT-4o participants for each task are available in the Supplementary Materials.

\subsection{Human performance} \label{appendix_human}\label{Human}
This section presents the interaction for the human participant S01.

\subsubsection{Verbal Production Tasks}

\paragraph{Verbal Fluency Test (VF)}
\paragraph{VF1 Instruction} 

\begin{quote}
    Now I will present you with a sheet of paper on which a word representing a category [animals] is written, and I will ask you to tell me as many words that come to your mind that belong to that particular category. You will have one minute. For example, for the category Trees, you might say: Oak, Poplar, Fir, etc.
\end{quote}

\paragraph{VF1 Response} 
\begin{quote}
Lion, Horse, Cow, Kid, Sheep, Pig, Dog, Cat, Mouse, Giraffe, Rhinoceros, Coyote, Eagle, Condor, Dolphin,  Whale, Hummingbird, Pigeon, Dove,  Rabbit,  Deer,  Roe Deer, Little Sparrow,  Fallow Deer, Panther, Cheetah, Leopard.
\end{quote}

\paragraph{VF2 Instruction} 
\begin{quote}
    
Now, the category is [fruit and vegetables]. You will have one minute.

\end{quote}

\paragraph{VF2 Response} 
\begin{quote}

Banana, Apple, Apricot, Strawberry, Blackberries, Raspberries, Blueberries, Lettuce, Valerian, Peppers, Potatoes, Onions, Zucchinis,  Shallot, Cauliflower,  Broccoli, Brussels Sprouts, Peach,  Mango, Papaya, Tomato, Avocado, Turnip, Nettle, Lemon, Orange.

\end{quote}

\paragraph{Free Association Task (FA)}
\paragraph{FA1 Instruction} 

\begin{quote}

Now I will present you with a sheet of paper on which
a word [building] will be written, and I will ask you to
tell me as many words as come to your mind that are
associated with the presented word. You will have one
minute. For example, for the word Father, you might
say: Mother, Son, Family, etc.
\end{quote}

\paragraph{FA1 Response} 

\begin{quote}
    Condominium, Building, Families, Parties, Windows, Elevator, Stairs, Doorbell, Rug, Entrance, Dog, Cat,  Cloister, Garden, Garage, Machine, Street, Pedestrian Crossing, Semaphore, Concierge, Doorman, Mail, Postbox, Neighbors, Tenants.
\end{quote}

\paragraph{ FA2 Instruction} 

\begin{quote}

Now, the word is [notebook]. You will have one minute.

\end{quote}

\paragraph{FA2 Response} 

\begin{quote}

Pencil, Quill, Eraser, Sharpener, Assignments,  Pencil Case,  Diary, Backpack, Italian, Mathematics, Squares, Colors, Drawing, Credibility, Ruler, Squad, Scissors, Teachers, Teachers, Companions, Games, Board, Plaster,  Explanations, Desks.				\end{quote}																				

\subsubsection{Alternate Uses Task (AUT)}
\paragraph{AUT1 Instruction}

\begin{quote}
    Now I will present you with a sheet of paper on which
a word [shoe] denoting an object is written. The task
consists of producing original uses for that object. All
objects have a common use. For example, what is the
common use of a tablecloth? To put it on the table when
we set it, in order to avoid getting it dirty. But that is the
common use; what we are interested in are the original
uses of the object, which I will tell you in a moment.
For instance, if we take the tablecloth off the table and
put it on ourselves, it can become a blanket if we are
cold or a cape if we are playing with a child. Or, if it is
old, we could cut it up and use it as rags. Your task is to
come up with as many original alternative uses for that
object as possible. You will have three minutes.
\end{quote}

\paragraph{AUT1 Response}
\begin{quote}

Place your feet in your hands to do a handstand; To get your hands dirty; Tie the shoes with laces that play a sort of game by holding the laces in your hand and rotating them; Glue many shoes together sideways forming a sort of wheel for playing;  Put in your ears when it's cold you could use the heel of some;  You can attach a pigment and use them to color;  You could take all the laces perhaps to make a necklace; Like bracelets.

\end{quote}

\paragraph{AUT 2 Instruction}
\begin{quote}

Now, the object is [fork]. You will have three minutes.

\end{quote}

\paragraph{AUT2 Response}

\begin{quote}
Like a decorative object when the wind blows many gold forks stick to it; To comb your hair; By tying two forks together with a rubber band you create a sort of chopstick; To take things and turn them around; In the garden with children you can use it to move earth or to create a soil for planting things; With clay it can be used to model to create figures;  As a musical instrument; By attaching cotton and placing fabrics you could create puppets or dolls;  When a branch becomes too crooked a garden can be used to straighten it;  To remove nails from walls;  To scrape off materials that can be scraped.

\end{quote}

\subsection{ChatGPT-4o performance} \label{AI}
This section presents the interaction for the ChatGPT-4o participant M43.

\subsubsection{Verbal Production Tasks}

\paragraph{Verbal Fluency Test (VF)}

\paragraph{VF1 Prompt}

\begin{quote}
Now I will present you with a word representing a category, and I will ask you to tell me as many words that come to your mind that belong to that particular category. You will have one minute. For example, for the category Trees, you might say: Oak, Poplar, Fir, etc. The word is "animals".
\end{quote}

\paragraph{VF1 Response}
\begin{quote}

Of course! Here is a list of animals that come to mind:

Dog, Cat, Lion, Tiger, Elephant, Giraffe, Rhinoceros, Hippopotamus, Deer, Wolf, Fox, Bear, Monkey, Dolphin, Shark, Whale, Penguin, Eagle, Owl, Snake, Crocodile, Turtle, Horse, Cow, Sheep, Goat, Hen, Duck, Swan, Crow, Wolf, Squirrel, Curly, Mole, Bat, Cheetah, Panther, Zebra, Ostrich, Ant, Bee, Wasp, Cicada, Kangaroo, Koala, Panda, Earthworm, Octopus, Jellyfish, Lobster, Shrimp, Seahorse, Gecko, Iguana, Parrot, Chameleon, Beaver, Ferret, Otter, Walrus, Elk, Platypus, Meerkat, Flamingo, Marmot, Carp, Tuna, Barracuda, Vulture, Coral, Cricket, Grasshopper, Falcon, Condor, Fly, Mosquito, Scarab, Tarantula, Scorpio, Snail, Snail, Frog, Toad, Salamander.

Time is up! I hope I have listed plenty!
\end{quote}

\paragraph{VF2 Prompt}
\begin{quote}
  Now, the category is "fruit and vegetables". You will have one minute.  
\end{quote}

\paragraph{VF2 Response} 
\begin{quote}
Here is a list of fruits and vegetables that come to mind:

Apple, Pear, Orange, Lemon, Mandarin, Grapefruit, Banana, Pineapple, Kiwi, Mango, Papaya, Coconut, Fig, Grape, Persimmon, Pomegranate, Fishing, Apricot, Cherry, Strawberry, Raspberry, Blackberry, Blueberry, Watermelon, Melon, Prune, Plum, Date, Avocado, Lychees, Currant, Pomegranate, Prickly Pear, Cranberry, Medlar, Star Fruit, Passion Fruit, Guava, Durian, Rambutan, Carrot, Potato, Onion, Garlic, Tomato, Pepper, Zuchini, Eggplant, Cucumber, Lettuce, Radish, Spinach, Chard, Cabbage, Cauliflower, Broccoli, Black Cabbage, Cabbage, Cabbage, Fennel, Celery, Parsley,
Basil, Rocket, Turnip, Beet, Asparagus, Mushrooms, Leek, Pumpkin, Green Bean, Broad Bean, Pea, Lentil, Chickpeas,
Bean, Soy, Horseradish, Radish, Chili, Endive, Jerusalem Artichoke.

Time is up! I hope I have listed plenty!
\end{quote}

\paragraph{Free Association Task (FA)}
\paragraph{FA1 Prompt} 

\begin{quote}
    
Now I will present you a word, and I will ask you to tell me as many words as come to your mind that are associated with the presented word. You will have one minute. For example, for the word Father, you might say: Mother, Son, Family, etc. The word is "building".
\end{quote}

\paragraph{FA1 Response} 

\begin{quote}
    Here is a list of words associated with "building":

Building, Skyscraper, Condominium, Apartment, Residence, Villa, Castle, Tower, Facade, Balcony, Terrace, Window, Door, Lobby, Entrance, Stairs, Elevator, Courtyard, Columns, Ceiling, Flooring, Roof, Attic, Basement, Cellar, Garage,
Concierge, Luxury, Stable, Ancient, Modern, Historical, Monument, Royal Palace, Fortress, Town Hall, Parliament,
Office, Room, Living Room, Room, Stay, Kitchen, Bath, Hallway, Library, Museum, Gallery, Inn, Hotel, Reception,
Decoration, Fresco, Sculpture, Stucco, Stained Glass Window, Marble, Bow, Capital, Pillar, Church, Basilica, Cathedral,
Cathedral, House, Abode, Quarter, Center, Metropolis, Street, Avenue, Square, Square.

Time is up!
\end{quote}

\paragraph{FA2 Prompt}  
\begin{quote}
Now, the word is "notebook". You will have one minute.
\end{quote}

\paragraph{FA2 Response}
\begin{quote}
Here is a list of words associated with "notebook":

Paper, Cover, Sheets, Pages, Lines, Squares, Notes, Writing, Pencil, Quill, Eraser, Highlighter, Ink, Diary, Agenda,
Notebook, Notepad, Register, School, School, Student, Teacher, Assignments, Theme, Drawing, Sketches, Graphs,
Rings, Tapping, Glue, Label, Title, Summary, Index, Margin, Paragraph, Calligraphy, Notes, Corrections, Notes, Study,
University, Library, Stationary Store, Binding, Form, Exams, Text, Underline, Write, To Note, Schemes, Concept Maps,
Summary, Organization, Planner, Draft, Revision, Archive, Teaching Material.

Time is up!
\end{quote}

\subsubsection{Alternate Uses Task (AUT)}

\paragraph{AUT1 Prompt}
\begin{quote}
    
Now I will present you with a word denoting an object is written.  The task consists of producing original uses for that object.  All objects have a common use. For example, what is the common use of a tablecloth? To put it on the table when we set it, in order to avoid getting it dirty.  But that is the common use; what we are interested in are the original uses of the object, which I will tell you in a moment.  For instance, if we take the tablecloth off the table and put it on ourselves, it can become a blanket if we are cold or a cape if we are playing with a child.  Or, if it is old, we could cut it up and use it as rags.  Your task is to come up with as many original alternative uses for that object as possible. You will have three minutes. The object is "shoe".
\end{quote}

\paragraph{AUT1 Response}
\begin{quote}
    
Here are some alternative and original uses of a shoe:
\begin{description}
    \item[1.] Plant pot - fill an old shoe with soil and plant flowers or herbs.
    \item[2.] Pen holder or pencil holder - perfect for an original and creative desk.
    \item[3.] Doorstop - a heavy shoe can prevent doors from slamming.
    \item[4.] Improvised musical instrument - filled with pebbles or rice, becomes a maracas.
    \item[5.] Phone holder - with the right angle, can hold a smartphone in a vertical position.
    \item[6.] Secret hiding place - perfect for storing small valuables inside an old shoe.
    \item[7.] Pet bowl - if cleaned well, can contain water or kibble for a cat or dog.
    \item[8.] Children's toy - with a little imagination it can become a toy car or a dollhouse.
    \item[9.] Artistic decoration - painted and hung on the wall, can be a modern sculpture.
    \item[10.] Mold for decorations - the sole can be used to create designs on painted surfaces or with clay.
    \item[11.] Insect trap - a perfect shelter for small insects, which can then be released outside.
    \item[12.] Water collector - in case of emergency, it can be used to collect rainwater.
    \item[13.] Weight for training - two shoes filled with sand can be used for physical exercises.
    \item[14.] Mini shield - to protect your hand in some emergency situation.
    \item[15.] DIY stamp - the sole can be carved to create a stamp with special patterns.
    \item[16.] Dog toy - an old shoe can become the perfect chew toy.
    \item[17.] Floating boat - if it is plastic and light, can be used for floating experiments.
    \item[18.] Substitute for a hammer - in the absence of a hammer, a stiff shoe can be used to nail something down.
    \item[19.] Improvised bag - tying the laces together, can carry small items.
    \item[20.] Cold hand cover - if it's very cold, you can put your hands inside to warm them.
    \item[21.] Knee protection - two shoes tied over the knees can serve as rudimentary knee pads.
    \item[22.] Mask for a costume - with holes it can become a helmet or a mask for a disguise.
    \item[23.] Mousetrap - with bait inside, could attract and trap small rodents.
    \item[24.] Sculpture for an art installation - a set of shoes can be used for a work of art.
    \item[25.] Doll raft - a floating shoe can carry a small toy on the water.
\end{description}

Time is up! I hope I have found creative ideas!
   \end{quote}

\paragraph{AUT2 Prompt}
\begin{quote}
    
Now, the object is "fork". You will have three minutes".

\end{quote}
\paragraph{AUT 2 Response}
\begin{quote}

Here are some alternative and original uses of a fork:
\begin{description}
    \item[1.] Phone holder - with the right fold, can hold a smartphone in place.
    \item[2.] Back scratcher - perfect for reaching difficult spots on the back.
    \item[3.] Paperweight - a heavy fork can hold papers in place on the desk.
    \item[4.] Photo holder - by bending the prongs, it can be used to hold up a photograph.
    \item[5.] Musical instrument - struck on a glass cup can produce harmonious sounds.
    \item[6.] Bag closure - inserted into an open bag, helps keep it sealed.
    \item[7.] Shoe brush - can be used to clean mud from the soles.
    \item[8.] Hard boiled egg holder - holds eggs in place for easy cutting.
    \item[9.] Mold for decorations - the prongs can leave textures on clay or salt dough.
    \item[10.] Improvised comb - useful for fixing hair in an emergency.
    \item[11.] Can opener - inserted under the lid of a can can make opening easier.
    \item[12.] Wall hook - fixed to the wall and folded, can be used to hang keys or towels.
    \item[13.] Gardening tool - perfect for digging small furrows in the earth.
    \item[14.] Place card at the table - inserted into a cardboard, can hold the names of the guests.
    \item[15.] Mini rake - useful for leveling the soil in pots.
    \item[16.] Hair accessory - inserted between the hair it can hold up a bun.
    \item[17.] Tea bag dripper - can help you squeeze the tea bags without burning yourself.
    \item[18.] Citrus peeler - the prongs can make it easier to cut the peel.
    \item[19.] Painting tool - great for creating texture on paint.
    \item[20.] Cake decorator - can be used to make designs on icing or sugar paste.
    \item[21.] Improvised skewer - perfect for skewering grilled foods.
    \item[22.] Ring holder - the prongs can hold rings and small jewelry in place.
    \item[23.] Ribbon curling tool - by passing a ribbon between the prongs, you can obtain a nice curl.
    \item[24.] Bookmark - a flat fork inserted between pages can act as a bookmark.
    \item[25.] Improvised weapon - in extreme situations, can be used for self-defense.
    \item[26.] Chandelier decoration - several forks can create lighting effects when hung.
    \item[27.] Easel for small paintings - folded in the right way, can support small canvases.
    \item[28.] Incense holder - perfect for inserting an incense stick and keeping it stable.
    \item[29.] Children's toy - with a little imagination, can become a character or a funny figure.
    \item[30.] Mini whisk - if folded, can be used to mix thick sauces.
\end{description}

Time is up! I hope I have found original ideas!
\end{quote}

\end{document}